\documentclass{aa}
\usepackage{graphicx}

\newcommand{\diff}{\mbox{${\rm d}$}}

\newcommand{\Teff}{\mbox{$T_{\rm eff}$}}

\newcommand{\comment}[1]{}
\newcommand{\beq}{\begin{equation}}
\newcommand{\eeq}{\end{equation}}
\newcommand{\beqa}{\begin{eqnarray}}
\newcommand{\eeqa}{\end{eqnarray}}
        \def\smallskip{\vskip 2pt}

\begin{document}


\title{Evolution of Planetary Nebulae I.\\
An improved synthetic model} 

\author{P. Marigo\inst{1,2} 
\and L. Girardi\inst{1,2} 
\and M.A.T. Groenewegen\inst{3,1}  
\and  A. Weiss\inst{1}}

\institute{
 Max-Planck-Institut f\"ur Astrophysik, Karl-Schwarzschild-Str.\
	1, D-85741 Garching bei M\"unchen, Germany \and
 Dipartimento di Astronomia, Universit\`a di Padova,
	Vicolo dell'Osservatorio 2, I-35122 Padova, Italy \and
 European Southern Observatory, Karl-Schwarzschild-Str.\
	2, D-85740 Garching bei M\"unchen, Germany 
}

\offprints{Paola Marigo, \\ \email{marigo@pd.astro.it}} 

\date{To appear in A\&A}

\abstract{
We present a new synthetic model to follow the evolution of a
planetary nebula (PN) and its central star, starting from the onset of
AGB phase up to the white dwarf cooling sequence.  The model suitably
combines various analytical prescriptions to account for different
(but inter-related) aspects of planetary nebulae, such as: the
dynamical evolution of the primary shell and surrounding ejecta, the
photoionisation of H and He by the central star, the nebular emission
of a few relevant optical lines (e.g. H$\beta$; He{\sc
\,ii}$\,\lambda$4686; [O{\sc \,iii}]$\,\lambda$5007).  Particular
effort has been put into the analytical description of dynamical
effects such as the three-winds interaction and the shell thickening
due to ionisation (i.e. the thin-shell approximation
is relaxed), that are nowadays considered important aspects of
the PN evolution.
Predictions of the synthetic model are tested by comparison with both
findings of hydrodynamical calculations, and observations of Galactic
PNe.  The sensitiveness of the results to the models parameters
(e.g. transition time, mass of the central star, H-/He-burning tracks,
etc.) is also discussed.  
We briefly illustrate the systematic differences that are
expected in the luminosities and lifetimes of PNe with either H- or
He-burning central stars, which result in different ``detection
probabilities'' across the H-R diagram, in both H$\beta$ and [O{\sc
iii}]\,$\lambda5007$ lines.
Adopting reasonable values of the model 
parameters, we are able to reproduce, in a satisfactory way, many
general properties of PNe, like the ionised mass--nebular radius 
relationship, the trends of a few  main nebular line ratios, 
and the observed ranges of
nebular shell thicknesses, electron densities, and expansion
velocities.  The models naturally predict also the possible 
transitions  from optically-thick to optically-thin configurations 
(and vice versa).
In this context, our analysis indicates that the 
condition of optical thinness to the H continuum plays an 
important role in producing the observed ``Zanstra discrepancy'' 
between the temperatures determined from H or He{\sc\,ii} lines, 
as well as it affects 
the mass-increasing part of the ionised mass-radius relation.
These predictions are supported by      
observational indications by M\'endez et al. (1992).
Another interesting result is that 
the change of slope in the electron density--nebular radius 
relation at $R_{\rm ion} \sim 0.1$ pc, pointed out by Phillips
(1998), is also displayed by the models and 
may be interpreted as the result of the progressive
convergence of the PNe to the condition of constant ionised mass.
Finally we would like to remark that, thanks to its computational
agility, our synthetic PN model is particularly suitable to
population synthesis studies, and it   
represents the basic ground from which many future
applications will be developed.
}

\maketitle

\section{Introduction}
\label{intro}

Planetary Nebulae (PN) are among the most beautiful and
fascinating celestial objects. They appear when 
low- and intermediate-mass stars, after losing their
external envelopes during the AGB phase, cross the H-R diagram 
on their way to the white dwarf 
cooling sequence. 
During this crossing, the central nuclei become 
sufficiently hot to ionise the circumstellar material
previously expelled.  
The resulting emission nebulae may turn bright 
enough (specially in the [O{\sc \,iii}] lines) to emerge strikingly  
against the background of ``normal'' stars, even at
relatively large distances. 

The study of the spectra of PNe 
provide unvaluable information on the chemical enrichment 
of the insterstellar medium from low- and intermediate-mass stars, 
and on the kinematics of their parent populations.
In general, they offer a unique tool to test stellar
evolution theory (e.g. mixing processes, stellar atmospheres,
evolutionary time-scales, etc.), as well the many physical processes
involved (e.g. stellar winds, gas dynamics and kinematics, 
photoionisation, line formation, etc.) 
The reader can get an overall picture by referring  
to the works of Pottasch (1984), Peimbert (1990), and Kwok (2000).

Indeed, the observed properties of PNe are the result of the 
complex interplay between several factors, whose
inter-relationship cannot be summarised in simple terms.
For instance, the PN lifetime -- i.e. the detectability period --
crucially depends on both i) the evolutionary time-scale of the central star
(the time it takes to become and remain an efficient
emitter of ionising photons) 
and ii) the dynamical time-scale 
of the circumstellar ejecta (including the duration of the ejection event
on the AGB tip, and the time taken by the nebula to expand away from the 
central star).

It is clear that the detectability of a PN requires a suitable tuning
between the two time-scales. If the central star evolves too quickly
the brightening of the ionised nebula 
would correspond to a short-lived event having a low detection 
probability. Conversely, if the central star 
evolves too slowly it would start to ionise the nebula   
when this has already dispersed to too large radii to be detected.
It follows that only when these two time-scales are comparable,
we can have a maximum probability of detecting the PN phase. 
As a matter of fact,   
neither the stellar evolutionary time-scale nor the dynamical time-scale
are simple predictable functions of the involved parameters.

Though all post-AGB stellar models indicate that, in general, 
the evolutionary speed on the H-R
diagram decreases at increasing final stellar mass, 
other factors play non-negligible roles (such as: past stellar history,
phase of the pulse cycle marking the termination of the AGB, 
efficiency of the fast wind, chemical
composition of the envelope; 
see Vassiliadis \& Wood 1994; Bl\"ocker 1995 for a detailed discussion).  
This complexity translates into quite large uncertainties
affecting, in particular, the so-called {\sl transition time} 
(that determines the onset  of the photoionisation phase), 
and the {\sl fading time} (that determines
the end of the PN lifetime). 
 
The dynamical time-scale of the nebula is also determined by several
factors including i) the properties (mass-loss rates and 
terminal velocities) of both the AGB (super- and/or slow) 
wind and the post-AGB fast wind, ii) their dynamical interaction producing   
gas shocks with related heating/cooling processes, and iii) the 
effect due to ionisation which tends to expand and accelerate the nebula as a 
consequence of the increased thermal pressure.
 
The importance of these dynamical factors has been pointed out 
by several past analytical works 
(e.g. Kahn 1983, 1989; Volk \& Kowk 1985; Kahn \& Breitschwerdt 1990; 
Breitschwerdt \& Kahn 1990), and then broadly confirmed and fully analysed by 
(magneto-)hydrodynamical simulations of PN evolution (e.g.  
Schmidt-Voigt \& K\"oppen 1987ab; Marten \& Sch\"onberner 1991; 
Balick et al. 1992; Mellema \& Frank 1995; Sch\"onberner \& Steffen 2000, and
references therein).

If, on one hand, hydrodynamical simulations are able to solve the fine details 
of the nebular structure getting insight into the involved physical processes, 
on the other hand they are so demanding in terms of computing time
that only few cases of PN evolution have so far
been adequately modelled.
The hydrodynamical approach does not permit 
a systematic study of PN properties as a function of the
fundamental  parameters yet,  i.e. stellar mass and metallicity, 
transition time, properties (mass-loss rates and terminal velocity) 
of stellar winds, etc. 
As a consequence, it becomes difficult also to predict the global 
properties of PNe populations in different galaxy environments 
-- e.g. how they should appear for varying histories of star 
formation and chemical enrichment. 

These aims may be achieved, taking a reasonable computational effort, 
with the aid of ``synthetic'' models of PNe, in which
the main factors determining their evolution are described 
by means of reasonable -- and necessarily simple -- approximations. 
Several examples of synthetic PNe models can be found in the 
literature, as in e.g. Kahn (1983, 1989), Kahn \& West (1985), 
Volk \& Kowk (1985), Stasi\'nska (1989), Kahn \& Breitschwerdt (1990), 
Breitschwerdt \& Kahn (1990), 
Stasi\'nska et al. (1998), Stanghellini \& Renzini (2000). 

However, it is worth making a due distinction within the class of 
synthetic models.
In fact, one can opt for either i) 
schemes based on rather crude assumptions 
(e.g. constant nebular mass and/or expansion velocity)
which do not attempt to predict the dynamical properties and/or the 
ionisation structure of PNe (e.g. Stanghellini \& Renzini 2000), 
or ii) more detailed analytical approaches which account for an overall, 
even if simplified, 
description  of PN evolution, including the dynamical one 
(e.g. Volk \& Kowk 1985; Kahn \& Breitschwerdt 1990).

As far as the dynamical aspect is concerned, a further distinction 
should be made. Most dynamical models for PNe 
(both synthetic and hydrodynamical) 
may  be  classified as either 
{\sl Three-Wind Models} if they consider the
possible interaction between the AGB slow wind, the AGB superwind,
and the post-AGB fast wind (e.g. Schmidt-Voigt \& K\"oppen 1987ab; 
Marten \& Sch\"onberner 1991), or 
{\sl Two-Wind Models} if they lack of a distinguishable
superwind phase (e.g. Kwok et al. 1978; Volk \& Kwok 1985), 
whose existence has actually  been supported by observational evidence
(see, for instance, van Loon et al. 1998). 

The present study belongs to the class of  
{\sl synthetic Three-Wind} models for PNe and
accounts, as far as possible, for many relevant aspects, including 
the dynamics of the nebula, the ionisation structure of H and He 
by adopting stellar fluxes from 
non-LTE model atmospheres, and the emission of a few
recombination and forbidden lines.
Of course, we cannot avoid to adopt some 
simplifying approximations and assumptions -- like
spherical symmetry, constant electron temperature. 
However, we will show that several tests (both theoretical and 
observational) support the overall performance of the model, 
so that it can be considered as a satisfactory tool for investigating
the general properties of PNe. 

The outline of the paper is as follows.
The basic ingredients of the model are 
fully described in Sect.~\ref{sec_pnmod}.
Then, after summarising the basic assumptions of the models, 
in Sect.~\ref{sec_overview} 
we give a brief overview of the model predictions.
We proceed in Sect.~\ref{sec_observ} with the exploration of the 
predicted properties 
of PNe as a function of the stellar mass, and of some free model parameters 
(e.g. transition time, electron temperature, H-/He-burning track).
To this aim, we consider several observables to be reproduced, always 
limiting to samples
of Galactic PNe. In this way, as a first step, we limit the
possible effects due to different metallicities, 
possibly present in PNe populations of different galaxies. 
Concluding remarks and future aims are expressed in Sect.~\ref{sec_con}.

\section{A new synthetic model of PN evolution}
\label{sec_pnmod}
\subsection{The general scheme}
\label{ssec_scheme}
 
Before proceeding with the detailed description of the model, it is
convenient to sketch the general scheme of
the several different processes here considered.  
	\begin{itemize}
	\item
We start with detailed synthetic models of AGB stars 
of varying initial mass and 
metallicity. According to the adopted prescription, mass loss 
gets more and more efficient as the star climbs the AGB, so that
most of the stellar envelope is ejected during the last ``superwind''
stages. We attribute proper expulsion velocities to the AGB ejecta, 
and follow their expansion during all the subsequent evolution.
	\item
The central star, peeled off of its envelope, is assumed to evolve
along the post-AGB track determined by its mass and 
initial metallicity. The so-called ``transition time'', corresponding 
to the very initial post-AGB evolution, is left as a free parameter.
At each time step, we compute the ``fast wind'' 
(momentum and kinetic energy) emitted by the central star,
as well as the flux of ionising photons. 
	\item
The fast wind shapes a shell nebula through its interaction 
with the slower stellar wind emitted during the AGB evolution.  
We compute the resulting 
dynamical evolution by means of the so-called 
``interacting-winds model''. As the shell expands, it continuously 
accretes matter from the circumstellar layers, ejected 
at earlier ages and with lower velocities during the AGB phase.
	\item
As soon as the flux of ionising photons becomes large enough (due to 
heating of the central star), an ionisation front proceeds through
the nebula. We follow its position as a function of time, as well
as the total emitted fluxes in a few relevant recombination and 
forbidden lines. 
	\item
The additional pressure provided by ionisation produces a 
thickening of the shell, that we follow in an approximative way.
	\item
Eventually, the nebula expands to very low densities, and the 
central star fades along its white-dwarf cooling sequence, causing both 
dispersion and dimming of the emission nebula. We follow the 
evolution of all nebular quantities up to sufficiently large times.
	\end{itemize}
The subsections below will illustrate in detail all the prescriptions we
have adopted.

\subsection{The AGB phase}
\label{ssec_agb}
The AGB phase is described by detailed synthetic models, covering the 
evolution of low- and intermediate-mass 
stars from the first thermal pulse up to the complete ejection
of the envelope (details can be found in 
Marigo 1998, 2001; Marigo et al. 1996, 1998, 1999).

From these models we derive the mass of the central star as a function 
of the initial mass and metallicity 
of the progenitor, the mean radial gradient in
density and chemical composition 
of the ejected material as a function of time (assuming spherical
symmetry and stationary wind flows).
A simple attempt to account for possible dynamical interaction 
of AGB shells ejected  at different epochs with different velocities
is presented  in Sect.~\ref{sssec_agbint}.

Mass-loss on the AGB is included according to the
semi-empirical formalism developed  by Vassiliadis \& Wood (1993;
hereinafter also VW93),
which combines observations of variable AGB stars 
(Mira and OH/IR
stars) with standard predictions of pulsation theory. 

In brief, mass-loss  during the AGB consists of two phases.
Initially, for $P < 500$ days, the mass-loss rate exponentially
increases with the pulsation period $P$
\begin{equation}
\log \dot M = -11.4 + 0.0123\,P 
\label{mlr1}
\end{equation} 
until when, for longer periods, it attains and maintains nearly
constant values, typical of the super-wind regime
\begin{equation}
\dot M =  6.07023 \times10^{-3} \beta \frac{L}{c \, V_{\rm AGB}}.
\label{mlr2}
\end{equation}
where $\beta = 1.13 \, (Z/0.008)$ following Bressan et al. (1998).
Equation (\ref{mlr2}) is derived simply assuming that all the momentum
carried by the stellar radiation field is transferred to the outer
gas, and allowing the possibility for 
multiple scattering of photons ($\beta \ge 1$; typically $\beta \sim 2$
for solar metallicity).
 
In Eqs.~(\ref{mlr1}) and (\ref{mlr2}), 
$\dot M$ is given in units of $M_{\odot}$ yr$^{-1}$, the stellar
luminosity $L$ is expressed in $L_{\odot}$, the pulsation period $P$ in
days,  the light speed $c$ and the 
terminal velocity of the
stellar wind $V_{\rm AGB}$ are given in km s$^{-1}$.

The pulsation period $P$ is derived from the period-mass-radius relation
(equation 4 in VW93), with the assumption that variable AGB stars are pulsating
in the fundamental mode (see Wood et al. 1999):
\begin{equation}
\log P = -2.07 + 1.94\, \log R - 0.9\, \log M
\label{periodo}
\end{equation}
where the period $P$ is given in days; the stellar radius $R$ and mass $M$ are
expressed in solar units.
The wind expansion velocity $V_{\rm AGB}$ (in km s$^{-1}$) is calculated 
as a function of the
pulsation period P (in days): 
\begin{equation}
V_{\rm AGB} = -13.5 + 0.056\, P\, .
\end{equation}
In VW93
the allowed variability range was constrained according to observations
of Mira and OH/IR stars.
Specifically, lower and upper limits of
$V_{\rm AGB}^{\rm min}  = 3 {\rm km\, s}^{-1}$ and 
$V_{\rm AGB}^{\rm max}  = 15 {\rm km\, s}^{-1}$,
respectively, were adopted.

At this point it should be remarked that, for our purposes,  
$V_{\rm AGB}^{\rm max}$ is a quite crucial quantity,
since it essentially determines the characteristic velocity 
of the superwind ejecta, from which most PNe are thought to originate.
Therefore, setting $V_{\rm AGB}^{\rm max}$ is practically equivalent to
specify the expansion velocity of the main ejection event at the end 
of the AGB phase,
which is clearly of great relevance to the subsequent dynamical 
evolution of the PN.
A few alternative 
prescriptions for $V_{\rm AGB}^{\rm max}$ are presented below.

\subsubsection{The maximum velocity}
\label{sssec_vagbm}
Several factors affect the terminal velocity of the AGB wind.
Observations of OH/IR and Mira variable stars indicate that 
$V_{\rm AGB}$ may range over a relatively large
range, i.e. from  $\sim 8$ to $30 $ km s$^{-1}$,
with a typical mean value of  $\sim 14-15$ km
s$^{-1}$  for galactic variables
(e.g. Gussie \& Taylor 1994; Groenewegen \& de Jong 1998;
Groenewegen et al. 1998, 1999; van Loon 2000),
and possibly lower for variables belonging to more metal
poor galaxies (e.g. 
$V_{\rm AGB}$(LMC)/$V_{\rm AGB}$(Galaxy) $\sim 0.5-0.6$ for OH/IR stars,
as pointed out by Wood et al. 1992).
In the framework of radiation-driven dusty winds, 
this fact would be the result of lower dust-to-gas ratio, $\delta$, 
at decreasing 
metallicity (e.g. $V_{\rm AGB}\propto \delta^{0.5}$ according to the 
theoretical expectations of Habing et al. 1994).
Moreover, both observations 
(e.g. Loup et al. 1993) and model predictions (e.g. 
Netzer \& Elitzur 1993) converge 
to the result that $V_{\rm AGB}$ is larger for carbon-rich stars
than for oxygen-rich stars, owing to the different absorption
properties of chemically different dust grains.

In addition to a dependence on chemical composition, calculations of
radiative transfer in dust-driven winds 
(e.g. Netzer \& Elitzur 1993; Habing et al. 1994; Kr\"uger et al. 1994;
Ivezic \& Elitzur 1995)
have pointed out that the 
terminal velocity is affected by several other factors, i.e.
mass-loss rate, radius of dust grain, and stellar luminosity.
 
A theoretical finding, which is relevant to the present analysis, 
is that  at sufficiently high mass-loss rates
($\dot M \ga 10^{-6}\, M_{\odot}$), the terminal velocity is expected to
be practically independent from $\dot M$ (Habing et al. 1994).

\begin{figure}
\centering
\resizebox{0.9\hsize}{!}{\includegraphics{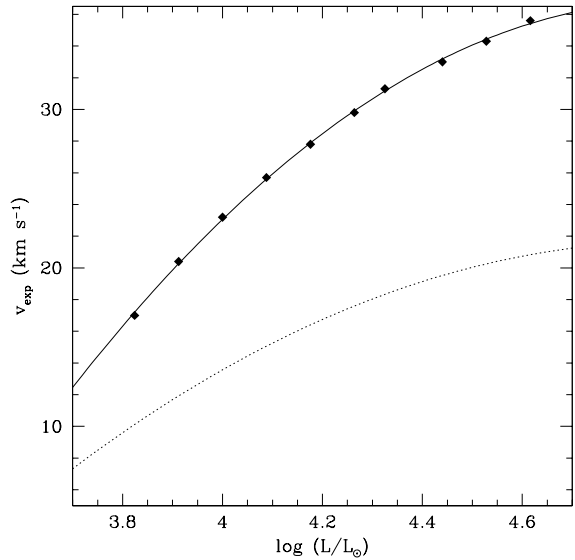}}
\caption{Expected dependence of $V_{\rm AGB}$ 
on stellar luminosity, according to Kr\"uger et al. 1994
(squares). The parabolic fitting relation (Eq.~\ref{eq_fitvexp}) 
is shown (solid line),
together with that obtained scaling it by a factor $f_{v} = 0.588$
(dotted line). See Sect. \ref{sssec_vagbm} for explanation}
\label{fig_vagbm}
\end{figure}
	
On the basis of all these aforementioned elements 
(both theoretical and observational), let us now consider some  
prescriptions we may reasonably adopt for $V_{\rm AGB}^{\rm
max}$, namely:
	\begin{enumerate}
\item $V_{\rm AGB}^{\rm max} = $ constant, say 15 km s$^{-1}$,
	regardless of any other parameter. This is the simplest 
	possible assumption, and  is made for instance,   
	by VW93 and Marigo et al. (1996, 1998).
\item  
	\begin{equation}
	V_{\rm AGB}^{\rm max} = f(Z) = 6.5\,(Z/0.008) + 0.00226\,P\, , 
	\end{equation}
	which includes a metallicity
	dependence, according to an empirical calibration 
        based on OH/IR variables of the 
	Galaxy and LMC (Bressan et al. 1998). 
        Such a prescription has been employed
	in the AGB synthetic calculations adopted here
	(Marigo et al. 1999; Marigo 2001).
	For typical values of $P > 500 - 2000$ days, we get
	$V_{\rm AGB}^{\rm max} \sim 8-11$ km s$^{-1}$ for $Z=0.008$, 
	and $V_{\rm AGB}^{\rm max} \sim 17-20$ km s$^{-1} $ for $Z=0.02$.
	We note that, at fixed metallicity, $V_{\rm AGB}^{\rm max}$ is almost
	constant, due to the weak dependence on the 
	pulsation period.
\item
	\begin{equation}
	V_{\rm AGB}^{\rm max} = f(L) = A\,(\log L)^2 + B\,\log L + C\,
	,
	\label{eq_fitvexp}
	\end{equation}
	which includes a luminosity dependence, according to 
	the results of two-fluid-model calculations for dust-driven
	winds (Kr\"uger et al. 1994). The fitting coefficients 
	(to reproduce their figure 12, see also Fig.~\ref{fig_vagbm}) are
	$A=-16.7282$, $B=164.1858$, and $C=-366.0314$. 
	Of course, these numbers only apply
	to a specific set of calculations performed by Kr\"uger et
	al. (1994).
	Therefore, in order to specify 
	$V_{\rm AGB}^{\rm max}$ in our model, 
	it may be reasonable to keep
	the relative trend with the stellar
	luminosity as predicted by Kr\"uger et al., 
	and scale it by a proper multiplicative factor, $f_{v}$,
	calibrated on some suitable constraint (Sect.~\ref{sssec_masrad}).
	\end{enumerate}
\subsection{The post-AGB evolution of the central star}
\label{sec_tracks}
The post-AGB tracks
are taken from Vassiliadis \& Wood (1994) data set, that
spans a large range of core masses  
($\sim$ 0.55 $M_{\odot}$ -- 0.95 $M_{\odot}$) and 
metallicities ($Z=$0.016, 0.008, 0.004, and 0.001). 
They also include both H- and He-burning central stars,
i.e. stars that have left the AGB at different phases of
their thermal pulse cycles, while they were efficiently burning 
their nuclear fuel either in the H- or in the He-burning
shell. These two situations lead to very different lifetimes
during the post-AGB, including the possible presence of a
late flash and ``final loop'' for He-burners.
All evolutionary tracks start at $\log T_{\rm eff} = 4$ which, in the
case of He-burning stars, marks the point in the 
blue-ward part of the last loop in the H-R diagram.
For all the details about the models 
the reader should refer to the Vassiliadis \& Wood (1994) paper.

\begin{figure}
\resizebox{\hsize}{!}{\includegraphics{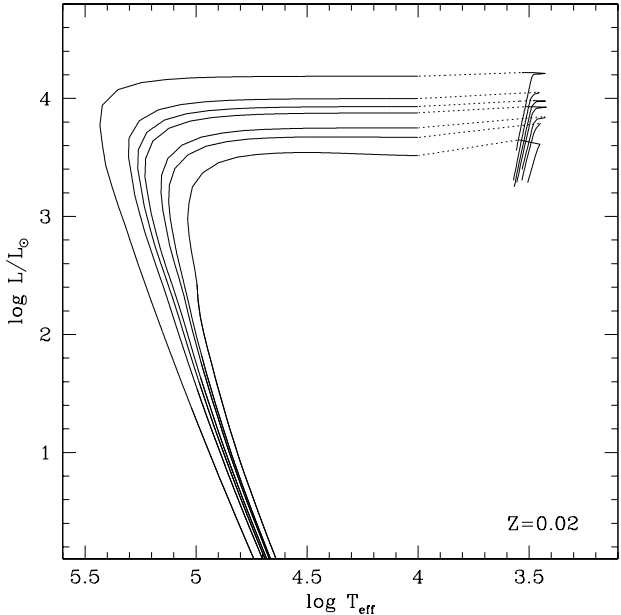}}
\caption{Evolutionary tracks in the H-R diagram from the first thermal
pulse on the AGB to the cooling sequences of  White Dwarfs.
The transition from the tip of synthetic AGB models to the
starting point of the post-AGB H-burning tracks of the same core  mass is shown
(dotted line)}
\label{fig_match}
\end{figure}

At fixed metallicity, the
match between the adopted AGB synthetic models and the post-AGB
tracks is performed by
interpolating the latter data as a function of the stellar mass
left at the end of the AGB. An example is illustrated in  
Fig.~\ref{fig_match}.

Minor differences between the stellar luminosity at the tip of the AGB
and that of the horizontal part of the post-AGB track in the H-R
diagram reflect small differences between the core mass-luminosity
relation derived by Vassiliadis \& Wood (1993, 1994), and that
adopted in the synthetic AGB models here in use 
(Wagenhuber \& Groenewegen 1998).

As we can see, the early post-AGB stages (i.e. after the star
has left the AGB up to $\log T_{\rm eff} = 4$) are not covered
by the tracks (dotted line in Fig.~\ref{fig_match}). 
They correspond to the transition from the end of the AGB to the starting
of the ionisation phase, i.e. the period in which the central star
does not emit enough UV photons to ionise the H atoms in the 
surrounding nebula.  
Large uncertainties affect the duration of such phase -- 
the so-called {\sl transition time}, $t_{\rm tr}$ (see
Renzini 1989; Sch\"onberner 1990) -- 
essentially due to our 
poor knowledge of the properties of the stellar wind, 
as it evolves from the super-wind regime to the fast-wind regime
(see Bl\"ocker 1995 for an analysis of this point).

Moreover, even for a given prescription for mass-loss, Vassiliadis 
\& Wood (1994) find no evident correlation between $t_{\rm tr}$ and 
the mass of the central star, 
but rather a dependence on the phase of the pulse
cycle at which the star leaves the AGB (Vassiliadis \& Wood 1994).
For all these reasons, we choose to adopt $t_{\rm tr}$ as a free parameter
in our model, comprised in the range 500 yr -- 10\,000 yr.   
%
\begin{figure*}[t]
\centering
\resizebox{0.7\hsize}{!}{\includegraphics{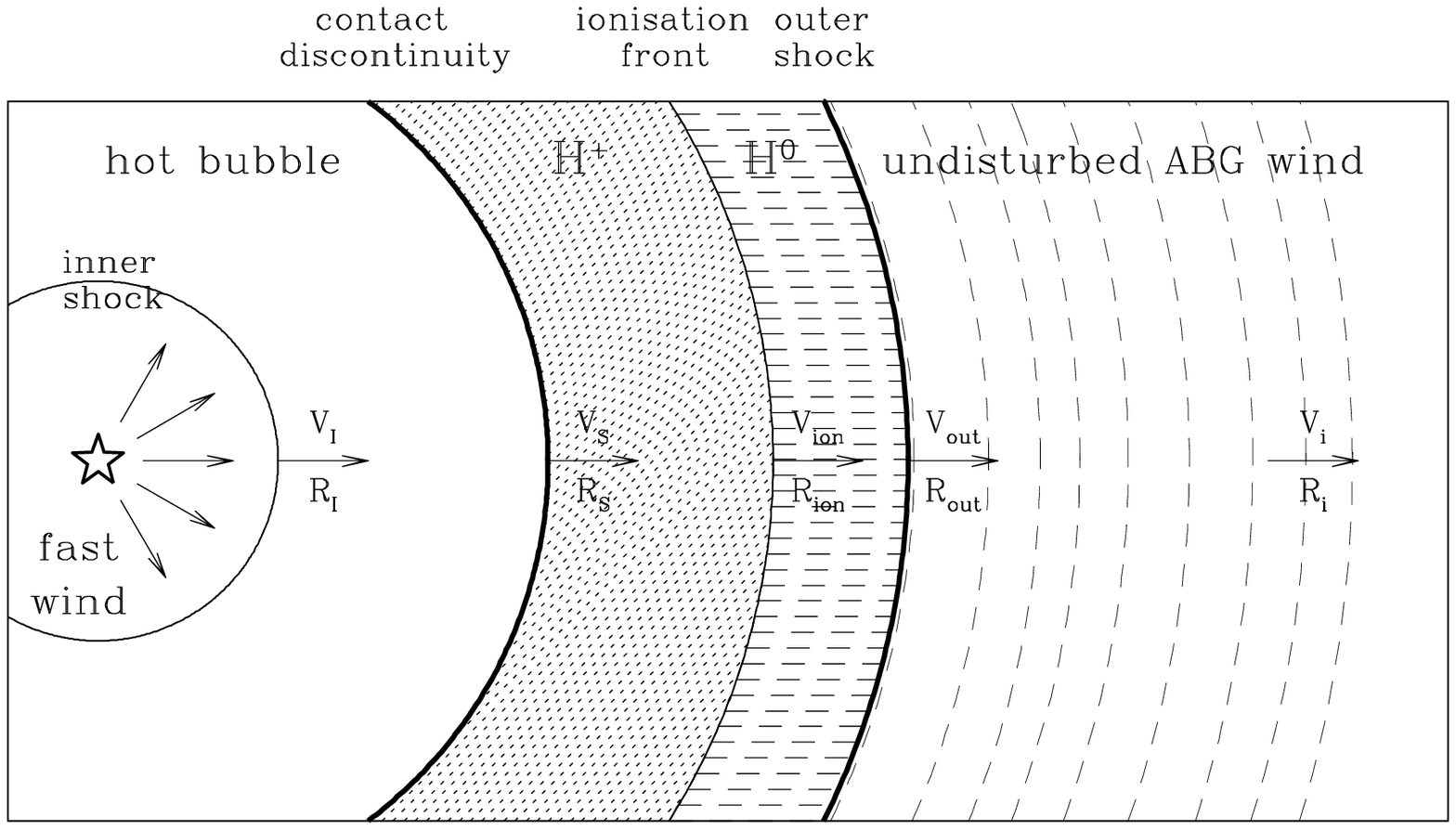}}
\caption{Schematic sketch of the system composed
by the central star, shocked PN shell, and undisturbed AGB wind}
\label{fig_pndisegno}
\end{figure*}
%
\subsubsection{The fast wind}
\label{sssec_fastw}
Mass-loss during the post-AGB phase of the central star
is a crucial process  as it affects the evolutionary time-scales and
hence observable properties of PNe.

In this model, we describe the fast wind adopting the prescriptions
from Vassiliadis \& Wood (1994).
This formalism combines the results
of the radiation-pressure-driven-wind theory for hot stars 
(Pauldrach et al. 1988), with empirical correlations observed 
for both AGB stars and central stars of planetary nebulae.
In this way, a continuous change  of the wind parameters is allowed,
as the evolution proceeds from the super-wind regime, 
typical of the last stages of the AGB
phase, to the accelerating fast wind regime during the post-AGB phase. 

Let us summarise the basic prescriptions.
The terminal wind velocity, $v$, of the fast wind is calculated with
\begin{equation}
\label{vterm}
\log \frac{v}{v_{\rm esc}} = -2.0 +0.52\,\log T_{\rm eff},
\end{equation} 
where the escape velocity is
\begin{equation}
v_{\rm esc} = \left [ (1-\Gamma) \frac{2 G M}{R} \right ]^{1/2}, 
\end{equation}
and $\Gamma$ is the ratio of the stellar to the Eddington
luminosity. This latter is given by
\begin{equation}
L_{\rm Edd} = \frac{0.2\,(1+X)}{4 \pi c\, G M}
\end{equation}
where the other symbols  have the usual meaning.
Equation~(\ref{vterm}) is a linear fit to observed data, including
central stars of PNe, B stars, A-F supergiants, and a fiducial point
for OH/IR stars.

Let us denote by $\dot M_{\rm lim}$ the radiation-pressure-driven
mass-loss limit, corresponding to the case in which all the momentum
carried by the radiation field is transferred to the wind gas:
\begin{equation}
\dot M_{\rm lim} = \frac{L}{c\,v}
\end{equation}
Then, the mass-loss rate, $\dot M$, in the PN phase is calculated assuming a
linear relation between 
$\log \dot M / \dot M_{\rm lim}$ and $\log T_{\rm eff}$: 
\begin{equation}
\log \frac{\dot M}{\dot M_{\rm lim}} = -3.92 + 0.67 \,\log T_{\rm eff}
\end{equation} 
as derived by Vassiliadis \& Wood (1994) following  
the theoretical results by Pauldrach et al. (1988).

\subsection{The dynamical evolution of the shocked nebula}
\label{ssec_dyn}
The dynamical evolution of planetary nebulae is followed in the
framework of the interacting-winds model, first developed by
Kwok et al. (1978), Kwok (1983), Kahn (1983), and substantially improved  
by Volk \& Kwok (1985, hereinafter also VK85).
Major improvements to the VK85 scheme are included, by
combining complementary prescriptions from other studies
(i.e. Kahn 1983; Kahn \& Breitschwerdt 1990; Breitschwerdt \& Kahn 1990),
as described in the next sections. 
 
For the sake of clarity, only the basic concepts and 
definitions are briefly recalled in the following 
(see also Table~\ref{terms}).
For more details the reader should refer to   
the aforementioned papers.

The dynamical evolution of PNe 
is considered  the result of the
interaction between the fast wind emitted by the central star during
the post-AGB evolution, and the slow circumstellar wind previously
ejected during the AGB phase. When the fast-moving gas catches up with
the slower one, a dense shell forms and two shock waves are generated.
The inner shock is located at the point where the fast wind encounters
the slow-moving gas. The outer shock is at the outer boundary of the
shell, beyond which the undisturbed AGB wind is expanding outward. 
Depending on the efficiency of radiative cooling, 
the hot shocked gas from the fast wind may fill (or not)
the bubble extending from the inner shock up to the so-called contact
discontinuity, i.e. the inner rim of the shell. No heat
or gas is assumed to pass across it.
The shocked shell may be partially or totally ionised by the
UV photons emitted by the central star.
 
\begin{table}
\caption{Adopted notation for relevant quantities}
\label{terms}
\begin{tabular}{lcl}
\noalign{\smallskip}
\hline
\noalign{\smallskip}
 $\dot M_{\rm AGB}$ & & mass-loss rate during the AGB phase\\
 $\dot m$ & & mass-loss rate during the post-AGB phase\\
 $M_{\rm S}$ & & mass of the shell  \\
 $R_{\rm I}$ & & radius of the inner shock  \\
 $R_{\rm S}$ & & radius of the contact discontinuity  (thin-shell)\\
 $R_{\rm S}^*$ & & radius of the contact discontinuity (thick-shell)\\
 $R_{\rm ion}$ & & radius of the ionisation front (thick-shell) \\
 $R_{\rm out}$ & & radius of the outer shock \\
 $V_{\rm AGB}$ & & terminal velocity of the AGB wind\\
 $v$  & & terminal velocity of the fast wind\\
 $V_{\rm I}$ & & velocity of the inner shock front \\
 $V_{\rm S}$ & & expansion velocity of the shell (thin-shell)\\
 $V_{\rm S}^*$ & & velocity related to $R_{\rm S^*}$ (thick-shell)\\
 $V_{\rm ion}$ & & velocity of the ionisation front (thick-shell) \\
 $P$ & & pressure of the hot shocked gas\\
\noalign{\smallskip}
\hline
\end{tabular}
\end{table}

According to the dominant mechanism causing the expansion of the
shell, we can distinguish two dynamical phases, namely:
the initial {\sl momentum-driven phase} and the subsequent 
{\sl energy-driven phase}.

Hereinafter we convene to set the evolutionary time
equal to zero ($t=0$) at the end of the AGB phase, i.e.
at the onset of the momentum-driven phase.

\begin{figure*}
\begin{minipage}{0.69\textwidth}
\resizebox{\hsize}{!}{\includegraphics{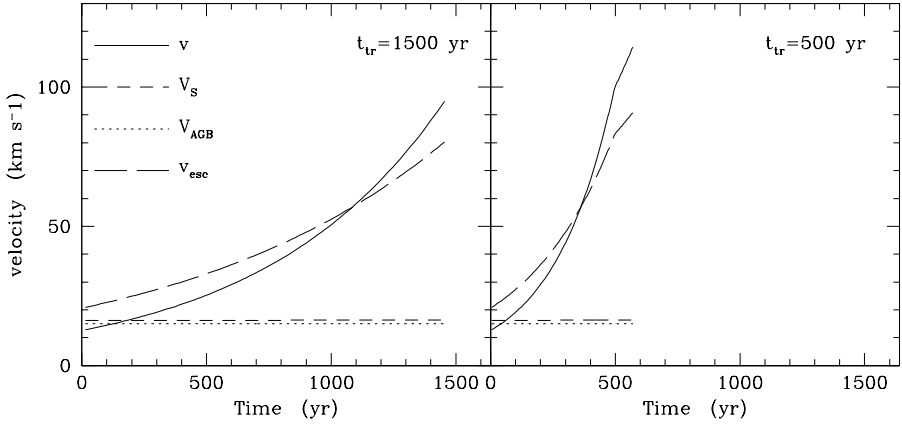}}
\end{minipage}
\hfill
\begin{minipage}{0.30\textwidth}
\caption{Relevant velocities during the momentum-driven phase
of the expanding shell surrounding a central star with mass 
0.6 $M_{\odot}$ and initial solar composition, 
for two choices of the transition time as indicated
left panel: model 1, right panel: model 5; see
Table~\protect{\ref{tab_mod}. Time is set to zero at the 
end of the AGB phase}
 }
\label{fig_mphase}
\end{minipage}
\end{figure*}

%
\subsubsection{The momentum-driven phase}
\label{momentum}
During the initial stages after the end of the AGB phase, 
the rate of radiative cooling is larger  than the rate of
kinetic input from the fast wind ($\dot m  v^{2}/2$), so that the inner
shock can be considered 
isothermal. The shocked matter can be compressed effectively and
the inner front is pushed quite close to the contact discontinuity
($R_{\rm I} \approx R_{\rm S}$), as it turns out from hydrodynamical
calculations (e.g. Marten \& Sch\"onberner 1991).
Under these conditions, the driving force which makes the shocked shell 
move outward is the transfer of momentum from the fast wind.
In our model, we account for this phase by adopting 
the prescriptions presented in Kahn \& Breitschwerdt (1990,
hereinafter also KB90).

{From} the conditions of mass and momentum conservation applied to  
the system composed by the fast wind and the swept shell, one gets 
simple relations expressing the time evolution of the shell mass
\begin{equation}
\frac{\diff M_{\rm S}}{\diff t} = \dot M_{\rm AGB}(t) 
\left[ \frac{V_{\rm S}(t)}{V_{\rm AGB}(t)} - 1\right]\, , 
\end{equation}
and the radius of the contact discontinuity
\begin{equation}
\frac{\diff R_{\rm S}}{\diff t} = V_{\rm S}(t) =  
\left[1+\sqrt{\alpha(t)}\right] V_{\rm AGB}(t), 
\end{equation}
where $\alpha$ is defined according to:
\begin{equation}
\label{eq_alpha}
\dot m(t) v(t) = \alpha(t) \dot M_{\rm AGB}(t) V_{\rm AGB}(t)
\end{equation}
For typical values of the winds parameters involved (e.g. $\dot
M_{\rm AGB} \sim 10^{-4} - 10^{-5} M_{\odot}$ yr$^{-1}$, $V_{\rm AGB}
\sim 10-20$ km s$^{-1}$, $\dot m \sim 10^{-7} - 10^{-8} M_{\odot}$
yr$^{-1}$, $v \sim 100 - 1000$ km s$^{-1}$) the quantity $\alpha$ is
found to be less than unity ($\sim 0.01-0.1$).
This means that the rate of momentum injection to the gas
from the accelerating fast wind is lower than that due to the 
AGB superwind, i.e. $\dot m v < \dot M_{\rm AGB} V_{\rm AGB}$.

Under the condition 
of a steady rate of momentum injection, i.e. if $\alpha =
{\rm const.}$, the shell's motion would be uniform, as 
assumed by KB90.
Here we consider the general case that 
$\alpha$ depends on time 
-- since all the quantities in
Eq.~(\ref{eq_alpha}) depend on time --, but find that 
$\alpha$ increases just negligibly in our models, thus recovering the
KB90 assumption that the velocity of the shell is almost constant 
during the momentum-driven phase.
An example is shown in  Fig.~\ref{fig_mphase} where 
$V_{\rm AGB} = 15.0$~km~s$^{-1}$ and  $V_{\rm S} \sim 16.3$~km~s$^{-1}$,
in very good agreement with the results of hydrodynamical calculations
by Kifonidis (1996).
The plot also shows the predicted speed-up of the fast wind 
(with velocity $v$ according to the 
prescriptions in Sect.~\ref{sssec_fastw}) and the escape velocity 
$v_{\rm esc}$ during the momentum driven phase.

The momentum-driven phase ends when the hot shocked  gas is able
to fill the interior in a time shorter than  the cooling time.
Following KB90, the change-over time $t_{*}$
from the momentum-driven phase to the energy-driven phase, crucially
depends on the velocity $v$ of the fast wind:
\begin{equation}
\label{eq_t*}
t_{*} = 5.23 \frac{\dot M_{\rm AGB}\, q \alpha}{(1+\sqrt{\alpha}) \,v^{6}}
\end{equation}
where $q=4\times 10^{32}$ cm$^{6}$ g$^{-1}$ s$^{-4}$ (Kahn 1976)
and all other quantities are in c.g.s. units.
We obtain that Eq.~(\ref{eq_t*}) is fulfilled 
typically for $v \sim 100 - 150$~km~s$^{-1}$, in agreement
with the findings of KB90. Clearly,  $t_{*}$ is quite 
sensitive to  the evolutionary speed of the central star during
the initial post-AGB stages, as illustrated  
in Fig.~\ref{fig_mphase} for two choices of the transition time. 

Finally, it is worth remarking that in all our models,  
during the momentum-driven phase,
the shell remains essentially neutral, the ionisation of 
H possibly starting towards the end of this phase and involving an
almost negligible part of the nebula.

\subsubsection{The energy-driven phase}
\label{energy}
The dynamical evolution of the shell at $t > t_{*}$ is mainly
controlled by the kinetic energy input from the fast wind, which
can be substantially thermalised as radiative cooling is inefficient.
A hot bubble develops, and the physical conditions of the interior
gas are those produced by a strong adiabatic shock. The front 
of the inner shock sits very close to the central star (i.e. $R_{\rm
I} \ll R_{\rm S}$).
During this phase the thermal pressure inside the hot bubble is   
the driving force acting on the shell. 

In our model, as soon as $t =  t_{*}$ we switch to the
dynamical description developed by VK85.
{From} the conditions of conservation of mass, momentum, and
energy applied to the system consisting of the hot bubble and the
shell, one finally ends up with a set of coupled, non-linear,
first-order differential equations:
\begin{eqnarray}
\frac{\diff R_{\rm S}}{\diff t} & = & V_{\rm S}
\label{eqr} \\
\frac{\diff M_{\rm S}}{\diff t} & = & \dot M_{\rm AGB} 
                         \left( \frac{V_{\rm S}}{V_{\rm AGB}} - 1 \right)
\label{eqm} \\
\frac{\diff V_{\rm S}}{\diff t} & = & \frac{1}{M_{\rm S}} 
			\left[4 \pi R_{\rm S}^{2} P - \dot M_{\rm AGB}
                        \frac{(V_{\rm S}-V_{\rm AGB})^{2}}{V_{\rm AGB}}\right]
\label{eqv} \\
\frac{\diff P}{\diff t} & = & \frac {(1/2) \dot m v^{2} - 10 \pi P_{\rm S} 
	V_{\rm S} R_{\rm S}^{2} + 
                 6 \pi P_{\rm S} V_{\rm I} R_{\rm I}^{2}}
		 {2 \pi (R_{\rm S}^{3}-R_{\rm I}^{3})}
\label{eqp}
\end{eqnarray}
Radiative cooling in the shocked  region is neglected, which is a
reasonable approximation during the energy-driven phase.
The numerical integration of Eqs.~(\ref{eqr}) -- (\ref{eqp}) is performed
explicitly, with a integration step much shorter (i.e. 1/100) than
the actual evolutionary time step. This latter is chosen so that the
maximum variation either in the effective temperature 
or in the luminosity of the
central stars cannot be larger than 0.01 dex.
The initial conditions for integration  are determined 
by the values of the variables
at the end of the momentum-driven phase.

It is worth remarking that in our model, both $\dot M_{\rm
AGB}$ and $V_{\rm AGB}$ are not constant as assumed in many past works,
but are given as a function of time according to the synthetic
calculations of the AGB phase (see Sect.~\ref{ssec_agb}). This means that, at each
instant, $\dot M_{\rm AGB}$ and $V_{\rm AGB}$ represent the mass-loss
rate and the expansion velocity of the wind layer previously ejected during
the AGB, which is {\it now} being swept by the shocked shell.

In Eq.~(\ref{eqp}) the radius ($R_{\rm I}$) and velocity ($V_{\rm I}$)
of the inner shock also enter. It was shown by 
hydrodynamical calculations (e.g. Marten \& Sch\"onberner 1991) 
that $R_{\rm I}$ rapidly decreases from its initial value 
($\sim R_{\rm S}$, as in the
momentum-driven phase) to a value small compared to $R_{\rm S}$.
Therefore, it is reasonable to adopt for $R_{\rm I}$ at each time step
the asymptotic solution as derived by VK85:
\begin{equation}
\frac{R_{\rm I}}{R_{\rm S}} \approx \sqrt
	\frac{3 \dot m v}{4 \dot M_{\rm AGB} V_{\rm AGB}}
		\left(\frac{V_{\rm S}}{V_{\rm AGB}} - 1 \right)^{-1}
\label{eqi}
\end{equation}
At each instant, the velocity $V_{\rm I} = \diff R_{\rm I} / \diff t$,
is derived from Eq.~(\ref{eqi}).

\subsubsection{Dynamical interaction of AGB shells}
\label{sssec_agbint}

As pointed out by hydrodynamical simulations 
(e.g. Frank et al. 1990, 1994; Steffen \& Sch\"onberner 2000;
Corradi et al. 2000), 
beyond the primary
shocked shell blown by the fast wind, dynamical interactions also occur 
in the ``undisturbed'' AGB wind, which may account for a number of 
complex morphological and emission features, 
i.e. crowns, edges, multiple shells, 
extended halos, etc.
It is obvious that a similar detailed analysis is out of reach
for our analytical model, nevertheless we can at least attempt
to give a zero-order description of AGB wind interactions.    

According to the adopted formalism for the AGB
wind (VW93), 
both $\dot M$ and $V_{\rm AGB}$ increase with time until
the natural onset of the superwind regime.    
In this scenario, it turns out that interactions between 
TP-AGB ejecta with different expansion velocities 
may be possible, even if they are not yet
reached by the outer shock front produced by the fast wind.

\begin{figure}
\centering
\resizebox{0.94\hsize}{!}{\includegraphics{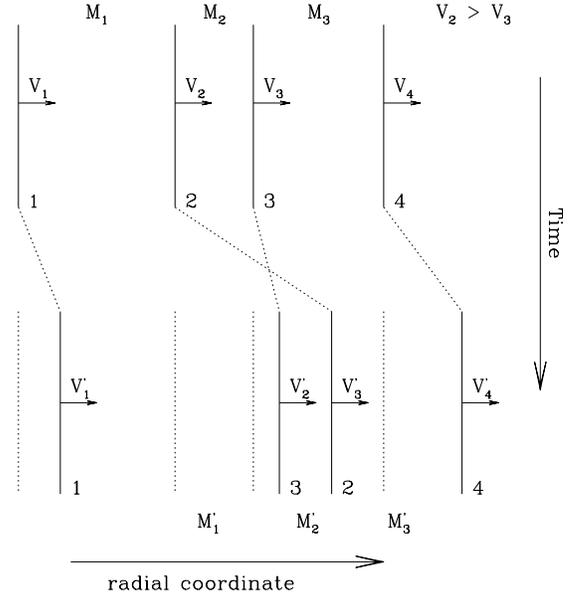}}
\caption{Schematic representation of the possible interaction
between expanding shells ejected during the AGB.
See text for a complete description}
\label{fig_overlap}
\end{figure}

Let us consider three consecutive AGB shells, with masses 
$M_1$, $M_2$, $M_3$, that are expanding with constant 
radial velocities at a given instant $t$. For the sake of simplicity, we  
denote with ($V_i$; $i=1,4$) the velocities at 
the shell boundaries of radii ($R_i$; $i=1,4$), so that
the $i^{\rm th}$-shell is moving with velocity $V_i$. 
The situation is sketched in Fig.~\ref{fig_overlap}.
Suppose also that $V_2 > V_3$.
As a consequence, it may happen that at a certain time $t+\diff t$ during
the expansion, shell 2 catches up shell 3. This would cause a sort
of overlap of gas shells with different dynamical properties, and
hence a change in the mass and velocity distributions  
along the radial coordinate.

To describe the final result 
(at time $t + \diff t$) of this interaction we adopt the following  simple
scheme. We denote with
\beq 
\label{eq_rp}
\left\{
\begin{array}{llll}
	R_1' =  R_1 + V_1 \, \diff t \\
	R_2' =  R_3 + V_3 \, \diff t \\
	R_3' =  R_2 + V_2 \, \diff t \\
	R_4' =  R_4 + V_4 \, \diff t 
\end{array}
\right.
\eeq
the new radial location of the shell boundaries (at increasing
distance; note the exchange between indices 2 and 3), and assume 
that $V_1'=V_1$ and $V_4'=V_4$ 
(i.e. at these radii the expansion motion is undisturbed).
Then, the problem is to determine
the five unknowns ($M_1'$, $M_2'$, $M_3'$, $V_2'$, and $V_3'$) 
which characterise at time $t+\diff t$ the shells delimited by radii
$R_j'$, as  given by Eqs.~(\ref{eq_rp}).
{From} imposing the mass conservation 
of the system composed by the three original shells, 
we get the new masses:
\beq 
\label{eq_mp}
\left\{
\begin{array}{lll}
M_1' = M_1 - \Delta M_1 \\
M_2' = M_2 + \Delta M_1 + \Delta M_2 \\
M_3' = M_3 - \Delta M_3 
\end{array}
\right.
\eeq
where 
	\beqa
	\Delta M_1 & = & \frac{{R_2'}^3 - {R_3'}^3}
	{{R_2'}^3 - {R_1'}^3}\, M_1 \\
\nonumber 
	\Delta M_2 & = & \frac{{R_2'}^3 - {R_3'}^3}
	{{R_4'}^3 - {R_3'}^3}\, M_3 \, ,
	\eeqa
whereas from the conditions of momentum and energy conservation:
	\beqa
	M_1 V_1 + M_2 V_2 + M_3 V_3 & = & M'_1 V'_1 + M'_2
	V'_2 + M'_3 V'_3 \\
\nonumber 
	M_1 {V_1}^2 + M_2 {V_2}^2 + M_3 {V_3}^2 & = & 
	M'_1 {V'_1}^2 + M'_2
	{V'_2}^2 + M'_3 {V'_3}^2 
	\eeqa
we obtain the new velocities $V_2'$ and $V_3'$
(for which we omit to present the formulas due to their trivial
derivation). 

\begin{figure}
\resizebox{\hsize}{!}{\includegraphics{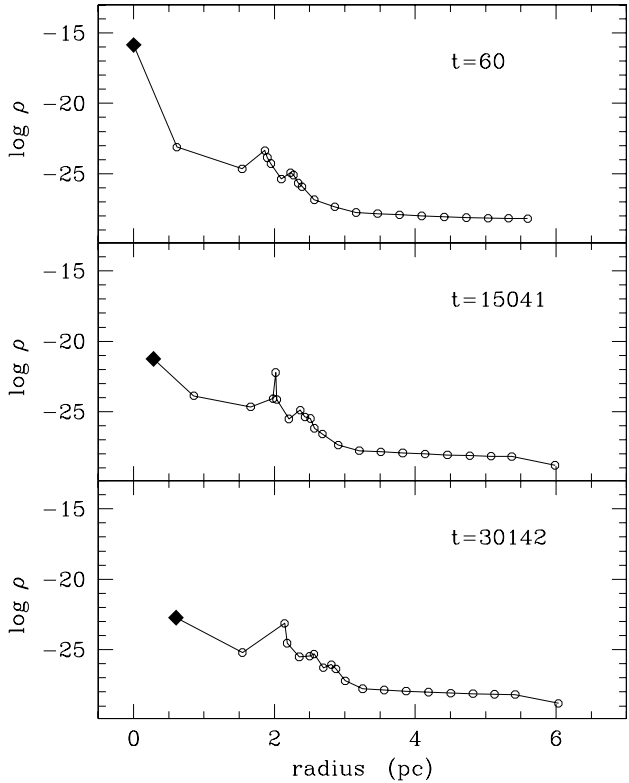}}
\caption{Evolution of the radial density (in gr cm$^{-3}$) structure 
of the circumstellar ejecta (model 4; see Table~\protect{\ref{tab_mod}}) 
at selected ages (indicated in yr) during the 
post-AGB phase.  $t=0$ would correspond to the moment in which
the central star leaves the AGB. 
In each panel, filled squares mark the average density in the primary
shocked shell, whereas open circles correspond to the sequence 
of unperturbed AGB shells}
\label{fig_densprof}
\end{figure}

This produces evident perturbations in the density pattern that would 
be otherwise expected for a stationary gas flow. 
A possible consequence of the shell overlap
sketched in Fig.~\ref{fig_overlap}
is the appearance of a 
density enhancement within the new shell 2 
(with mass  $M_2 + \Delta M_1 + \Delta M_2$).

According to the above prescriptions we find that this kind of dynamical
events begins to affect the circumstellar ejecta in the last stages 
of the AGB evolution (i.e. after the onset of the super-wind regime), and
usually involves shells ejected during the transition from the slow AGB
wind to the superwind. This feature is in agreement with
the hydrodynamical findings by Steffen \& Sch\"onberner (2000).
 
An example is shown in Fig.~\ref{fig_densprof}. We can notice that 
non-negligible 
density enhancements show up in the radial density structure of the model.
These density spikes are rather close to the primary shocked shell, while
the long tail of slow low-density shells remains unperturbed.

We are completely aware that such approach is an extremely
rough description of the real situation, which can be handled 
correctly only with the aid of hydrodynamics.
Nevertheless,
it may be interesting to get a first hint of likely interaction events 
(i.e. when and where they might occur) in the unperturbed AGB ejecta
surrounding the primary shocked nebula. 
Furthermore, we remark that
the dynamical scheme presented in this 
section does not affect any of the relevant predictions, 
as long as the PN is  assimilated to the primary shocked shell, as in the
analytical interacting-winds model (see Sect.~\ref{ssec_dyn}).

\subsection{Photoionisation rates for H and He$^{+}$}
\label{sec_photrates}
As the central star evolves at increasing effective temperatures along
the horizontal part of the H-R track, it starts emitting  UV photons that can
ionise the surrounding nebula. 
  
\begin{figure}
\resizebox{\hsize}{!}{\includegraphics{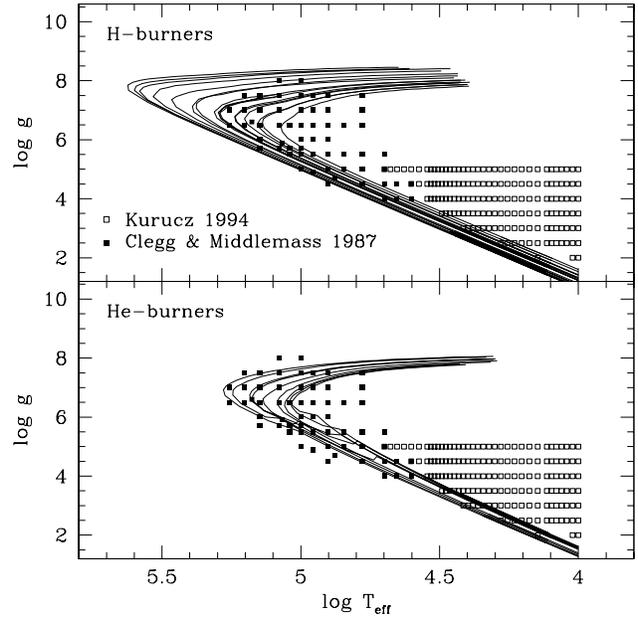}}
\caption{Available data grid 
for the ionising rates of H and He$^{+}$ 
in the $\log T_{\rm eff} - \log g$ plane according to model
atmospheres of  Clegg \& Middlemass (1987; filled squares) 
and Kurucz (1994; empty squares). 
The sets of H- and He-burning post-AGB
tracks from Vassiliadis \& Wood (1994) are superimposed}
\label{q_gtef}
\end{figure}

In this study we account for the 
photoionisation of H and He$^{+}$, which occurs
when a significant fraction of the emergent photons has
a wavelength shorter than 911~\AA\ 
(for $T_{\rm eff} \ga 30\,000$ K)
and 228~\AA\ (for $T_{\rm eff} \ga 60\,000\,{\rm K}$), respectively.
Ionisation fluxes (in cm$^{-2}$ s$^{-1}$) are
derived  from  atmosphere models, namely:
the LTE models calculated by Kurucz (1994)  in the 
ranges:  $10\,000 \le T_{\rm eff} \le 40\,000\,{\rm K}$, 
$2.0 \le \log g \le 5.0$; and the non-LTE models by   
Clegg \& Middlemass (1987) in the ranges: 
$40\,000\,{\rm K} \le T_{\rm eff} \le 180\,000\,{\rm K}$, 
$4.0 \le \log g \le 8.0$. In the adopted model atmospheres 
the abundance ratio He/H$ = 0.1$ (by number) is assumed. 

In Fig.~\ref{q_gtef} the adopted grids of model atmospheres 
are shown in the $\log T_{\rm eff} - \log g$ plane, together with 
the sets of post-AGB tracks for both H- and He-burner central stars.
For each grid point, ionising fluxes are obtained    
integrating the synthetic spectra 
	\beq
	Q = \int_{h \nu_1}^{h \nu_2}\frac{L_{\nu}}{h\nu} {\rm d}(h\nu)
	\label{eq_qion}
	\eeq
with $h \nu_1 = 13.6$ eV and $h \nu_2 = 54.4$ eV for  the H continuum,
and $h \nu_1 = 54.4$ eV and $h \nu_2 = \infty$ for  the He$^{+}$
continuum, thus neglecting absorption by neutral helium in
the nebula.
These calculations yield $Q({\rm H}^0)$ and $Q({\rm He}^+)$ as a function of
both effective temperature 
(in the range $T_{\rm eff}: 10-180\times 10^3$ K) 
and surface gravity (in the range $\log g: 2.0-8.0$) of the hot star. 

Within the $g-T_{\rm eff}$ range 
spanned by the model atmospheres, a
 bi-linear interpolation in $\log T_{\rm eff}$ and $\log g$ 
over this data grid is then adopted to evaluate the
current ionisation fluxes along the post-AGB tracks.
For possible points not covered by the grid,
we extrapolate the model data in $\log g$  for $T_{\rm eff} < 180\,000$ K, 
and we assume a blackbody spectral energy distribution for 
$T_{\rm eff} > 180\,000$ K 
(temperatures attained only by H-burning central stars with 
$M \ga 0.7\, M_{\odot}$; see Fig.~\ref{q_gtef}).

\begin{figure}
\resizebox{\hsize}{!}{\includegraphics{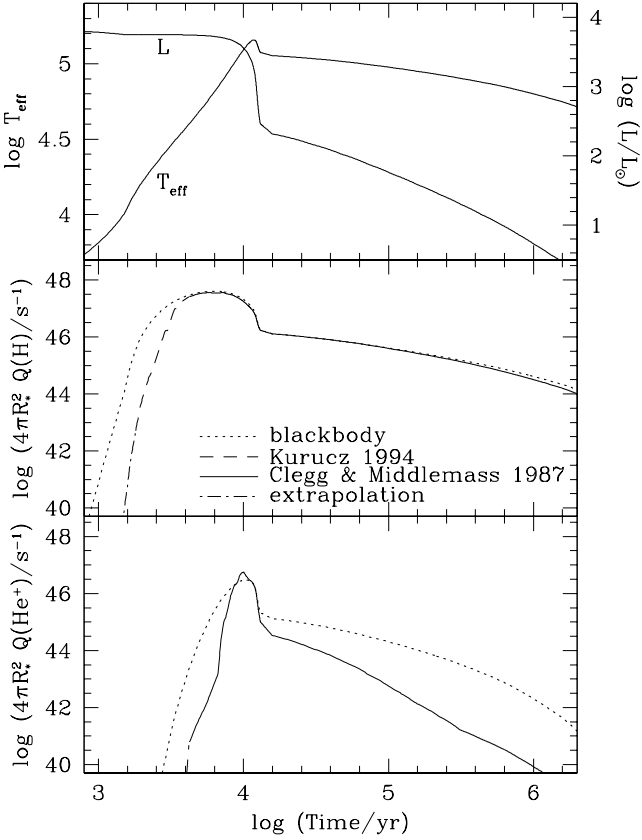}}
\caption{Ionisation rates as a function of time for a 0.6 $M_{\odot}$
post-AGB central star (model 4 in Table~\protect{\ref{tab_mod}}).
See text for more details}
\label{fig_ionf}
\end{figure}

An example is shown in Fig.~\ref{fig_ionf}, where the ionisation rates
for H and He$^{+}$ (i.e. $4 \pi R_{*}^2 Q$(H) and 
$4 \pi R_{*}^2 Q$(He$^{+})$ in s$^{-1}$, where $R_{*}$ is the stellar
radius) are calculated as a function of time for 
an evolving central star with a
mass of about 0.6 $M_{\odot}$. Predictions from 
Kurucz and Clegg \& Middlemass model
atmospheres  are compared to blackbody integrations. 

The temporal behaviour of the ionisation rates is controlled by the
changes in $T_{\rm eff}$ and $L$, and the evolutionary speed of the 
central stars. 
In Fig.~\ref{fig_ionf}
we can see that the photoionisation rates significantly increase
as the star heats up at nearly constant luminosity, reach a peak
(in proximity of the knee of the H-R track, at the maximum of 
$T_{\rm eff}$ in the top panel)
and then  quickly fall down when the luminosity drops.
Subsequently, the slower decrease of the ionisation rates reflects
the  drastic deceleration of the evolutionary speed.
We can also notice that in general $Q$(H) $>$ $Q$(He$^{+}$), and 
the peak duration is larger for $Q$(H) compared to $Q$(He$^{+}$).

As far as the ionisation of H is concerned, 
we notice that the blackbody assumption  
notably overestimates the continuum flux for $T_{\rm eff} < 40\,000$ K,
whereas it approximates well the atmosphere models 
at higher temperatures.

Much larger differences exist  between the He$^{+}$ ionisation fluxes
predicted according to the blackbody and stellar atmospheres models.
These discrepancies are due to important non-LTE effects 
that significantly alter the relative number of neutral, single, 
and doubly ionised He expected under LTE conditions (Clegg \& Middlemass
1987).

As shown in the bottom panel of Fig.~\ref{fig_ionf}, 
for most of the post-AGB evolution the adoption of a blackbody spectrum 
leads also to overestimate the He$^{+}$ continuum,
except for the emission peak corresponding to the highest effective
temperatures, before the decline of the stellar luminosity (see top panel).
Such remarkable excess of UV photons may be one of the main factors
at the origin of the so-called Zanstra discrepancy (refer to 
Gruenwald \& Viegas 2000 for a recent analysis; 
see also Sect.~\ref{ssec_zan}).

\subsection{Ionised mass}
\label{ssec_ionmas}
The ionisation structure of the nebula is solved with the aid of
simple but reasonable prescriptions.
Once the ionising rates  emitted by the central star are known, 
the ionised mass can be singled out, provided that
the current density stratification of the circumstellar
gas (i.e. shocked shell and unperturbed AGB shells) is determined.
The density of the ionised part of the shell is estimated from the equation 
of state of a perfect gas, assuming a constant electron temperature
$T_{\rm e}$ (typically  $~ 10\,000$ K).

Then, the ionised masses  for both H and He$^{+}$ are  derived 
assuming equilibrium between ionisation and recombination events.
It means, that within a certain volume, limited by the Str\"omgren 
radius $R_{\rm ion}$, the
rate of recombinations to all levels but the ground state, 
is equal to the number of ionising photons from the central star.
This condition translates into 
\beq
\label{eq_qh}
Q({\rm H}^0)  = 4 \pi 
        \int_{R^{*}}^{R_{\rm ion}({\rm H}^+)} \alpha_{\rm B}({\rm H}^0) 
	r^2 N_{\rm e} N({\rm H}^{+}) \diff r  
\eeq
for neutral hydrogen and 
\beq
\label{eq_qhep}
Q({\rm He}^+)  = 4 \pi  
        \int_{R^{*}}^{R_{\rm ion}({\rm He}^{++})} \alpha_{\rm B}({\rm He}^{+})
	r^2 N_{\rm e} N({\rm He}^{++}) \diff r  
\eeq
for single-ionised helium.
It is clear that these equations may be fulfilled  
as long as $R_{\rm ion} \le R_{\rm neb}$, where $R_{\rm neb}$
is the outermost radius of the gaseous nebula. 
The quantities
$\alpha_{\rm B}$  are the effective recombination coefficients,
expressing the total recombination rates  
(cm$^3$ s$^{-1}$) to all excited levels. They depend weakly on
the electron temperature $T_{\rm e}$ and can be expressed as
\beqa 
\alpha_{\rm B}(\rm{H}^0) & = & 
	2.54\, 10^{-13}\,\, (T_{\rm e}/10^{4})^{-0.791}\\
\nonumber
\alpha_{\rm B}(\rm{He}^{+}) & = & 
	1.52\, 10^{-12}\,\, (T_{\rm e}/10^{4})^{-0.716}.
\eeqa
according to the fitting formulas presented by Oliva \& Panagia (1983).
	
We assume that the ionisation radius for single-ionised helium
coincides with that of ionised hydrogen,
i.e. $R_{\rm ion}({\rm He}^{+}) = R_{\rm ion}({\rm H}^{+})$, 
which is a good approximation for $T \ga 4 \times 10^4$~K
(Osterbrock 1974).
Then, denoting by $N({\rm H})$ and $N({\rm He})$
the total number densities (neutral atoms $+$ ions;  in cm$^{-3}$) 
of hydrogen and helium, and by $N_{\rm e}$ the number density of
free electrons, we adopt the following scheme for the ionisation
structure (with obvious meaning of the particle number densities $N$):
\beq
\label{eq_ion1}
\left\{
\begin{array}{llll}
	 N({\rm H}^0)   =  N({\rm He}^0) = N({\rm He}^+)  =  0 \\
	\nonumber
	 N({\rm H}^+)   =  N({\rm H}) \\ 
	\nonumber
	 N({\rm He}^{++})  =  N({\rm He}) \\ 
	\nonumber
	 N_{\rm e}  =  N({\rm H}) + 2 \,N({\rm He}) 	
\end{array}
\right.
\eeq
for $r < R_{\rm ion}({\rm He}^{++})$;
%
\beq
\label{eq_ion2}
\left\{
\begin{array}{llll}
	N({\rm H}^0)  =  N({\rm He}^0) = N({\rm He}^{++}) =  0 \\
	\nonumber
	N({\rm H}^+)  =  N({\rm H}) \\ 
	\nonumber
	N({\rm He}^{+})  =  N({\rm He}) \\ 
	\nonumber
	N_{\rm e}  =  N({\rm H}) + N({\rm He}) 	
\end{array}
\right.
\eeq
for $R_{\rm ion}({\rm He}^{++}) \le r < R_{\rm ion}({\rm H}^{+})$, and
\beq
\label{eq_ion3}
\left\{
\begin{array}{llll}
	N({\rm H}^0)  =  N({\rm H}) \\ 
	\nonumber
 N({\rm H}^+)  =  N({\rm He}^+) = N({\rm He}^{++}) =  0 \\
	\nonumber
	N({\rm He}^{0})  =   N({\rm He}) \\ 
	\nonumber
	N_{\rm e}  =  0 	
\end{array}
\right.
\eeq
for $r > R_{\rm ion}({\rm H}^{+})$.

\subsection{Shell thickening due to ionisation}
\label{ssec_thick}
%
\begin{figure*}
\begin{minipage}{0.69\textwidth}
\resizebox{\hsize}{!}{\includegraphics{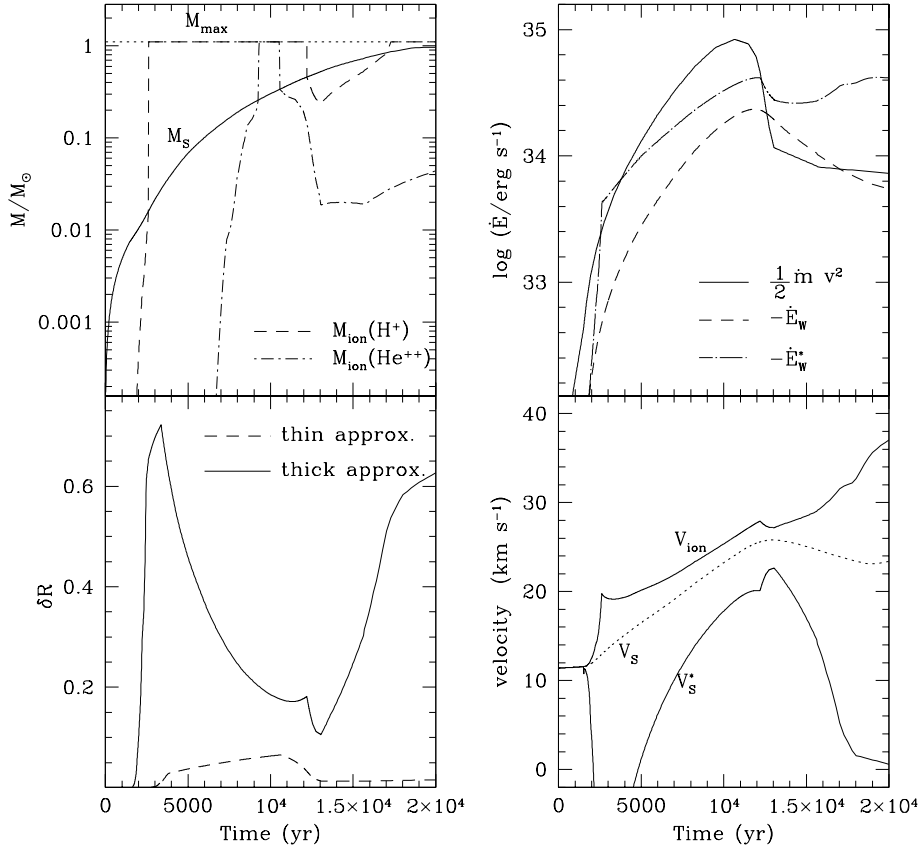}}
\end{minipage}
\hfill
\begin{minipage}{0.30\textwidth}
\caption{Predicted properties of the shocked nebula as a function 
of time, accounting for the dynamical effect resulting from 
ionisation (model 4 in Table~\protect{\ref{tab_mod}}).
{\bf Top-left panel:} Temporal behaviour of the 
mass of the shocked shell $M_{\rm S}$, and 
ionised masses for H$^{+}$ and He$^{++}$. 
The total mass $M_{\rm max}$ of the circumstellar AGB ejecta is also 
shown (dotted line).
{\bf Top-right panel:} Comparison between the rate of injection of
mechanical energy from the fast wind ($\dot m v^2/2$), and
the rate of the work done by the pressure in expanding the shell 
($- \dot E_{\rm W}$, $- \dot E_{\rm W}^*$; see text). 
{\bf Bottom-left panel:} Relative thickness $\delta R$ of the ionised
part of the shocked shell.
The prediction under the thin-shell
approximation is compared to the
case in which the shell thickening due to ionisation is 
included.
{\bf Bottom-right panel:} Velocities at the shell boundaries: 
$V_{\rm S}$ of the contact discontinuity under the thin-shell
approximation (dotted line);
$V_{\rm S}^*$ of the actual contact discontinuity, 
and $V_{\rm ion}$ of the ionisation front (solid lines) when the 
shell-thickening corrections are included. See text for more
details}
\label{fig_sthick}
\end{minipage}
\end{figure*}
An implicit assumption of the interacting-winds model 
(outlined in Sects.~\ref{momentum} and \ref{energy}) is that  
the shocked shell is thin compared to its radial extent, i.e. 
$\Delta R_{\rm S}/R_{\rm S} \ll 1$. 
However, observations indicate that ring-like nebulae 
exhibit a large range of relative thickness 
(from about 0.2 to 0.8, with the peak of the 
distribution at $\sim 0.4 - 0.5$), 
and even homogeneous nebulae lacking a central hole exist
(Zhang \& Kwok 1998).
Moreover, the expansion velocities of many young PNe can be 
notably high, up to $\sim 50\, {\rm km\, s}^{-1}$ (e.g. Weinberger 1989),
which is something not predicted in the framework of the
simple interacting-winds model.
 
Hydrodynamical models of planetary nebulae (Marten \&
Sch\"onberner 1991) have pointed out that these observations 
can be explained considering the dynamical effects
caused  by the ionisation of the nebula. 
In fact, thanks to the growing number of free electrons 
the thermal pressure increases inside the ionised nebula,  
which then tends to expand both inward and outward.
As a consequence, the inner and outer
nebular rims are decelerated and accelerated, respectively.

Such dynamical effect is included in this model
following the analytical treatment suggested by Kahn (1983).
The basic steps are shortly recalled below.
In practice, 
the effect of ionisation 
is treated as a linear perturbation of the
equations of momentum and energy conservation 
(Eqs.~\ref{eqv}--\ref{eqp}), by introducing a term
$\Delta P_{\rm ion}$ for the extra pressure.
Once the unperturbed solution  of  
the set of Eqs.~(\ref{eqr}) -- (\ref{eqp}) is singled out 
(in the thin-shell approximation), 
proper first-order corrections are then applied 
to the radial coordinate $R_{\rm S}$ and velocity $V_{\rm S}$ of the  
contact discontinuity according to the following scheme.


Let us assume that, as a consequence of the pressure contribution due to
ionisation, the neutral shell 
expands outward to a radius $R_{\rm S} +\Delta R_{\rm a}$ 
(with $\Delta R_{\rm a} > 0$), 
and the contact discontinuity expands inward down to a
radius $R_{\rm S} + \Delta R_{\rm b}$ (with $\Delta R_{\rm b} < 0$), 
where $R_{\rm S}$ refers to the
unperturbed value.
Then, the unperturbed energy equation
\begin{equation}
\frac{\diff}{\diff t} (2 \pi R_{\rm S}^{3} P) = \frac{1}{2} \dot m v^2
             -4 \pi R_{\rm S}^{2} P \frac{\diff R_{\rm S}}{\diff t}
\end{equation}
becomes 
\begin{equation}
\left.\begin{array}{l}
\displaystyle{
\frac{\diff}{\diff t} [2 \pi (R_{\rm S}+\Delta R_{\rm b})^{3} 
	(P+\Delta P_{\rm ion})]=} \\
\displaystyle{\frac{1}{2} \dot m v^2 
             -4 \pi (R_{\rm S}+\Delta R_{\rm b})^{2} (P+\Delta P_{\rm ion}) 
             \frac{\diff (R_{\rm S}+\Delta R_{\rm b})}{\diff t}} 
\end{array}
\right.
\label{eq_pert}
\end{equation}
An analogous treatment is applied to the unperturbed momentum equation.
The relationship between $R_{\rm a}$ and $R_{\rm b}$ is given by
equation 31 in Kahn (1983) to whom we refer for all details.
Expliciting the time derivatives and neglecting the second-order
terms, one eventually derives the radius 
corrections ($\Delta R_{\rm a}$, $\Delta R_{\rm b}$), 
the velocity corrections  
($\dot\Delta R_{\rm a}$, $\dot\Delta R_{\rm b}$), and the pressure 
corrections $\Delta P_{\rm ion}$, $\dot{\Delta P_{\rm ion}}$ 
as a function of the unperturbed quantities, winds parameters 
and ionised mass.


For the sake of clarity, we can now summarise our prescriptions. 
As the ionisation front breaks into the shocked shell, this latter 
will be characterised by three relevant radii, namely:
\begin{equation}
R_{\rm S}^{*} = R_{\rm S} + \Delta R_{\rm b} \,\,\,\mbox{: actual contact
discontinuity}
\end{equation}
\begin{equation}
R_{\rm ion} = R_{\rm S} + \Delta R_{\rm a} \,\,\,\,\mbox{: ionisation front}
\end{equation}
\begin{equation}
R_{\rm out} = \left[ R_{\rm ion}^{3} + \frac{3 M_{\rm neut}}
{4 \pi \rho_{\rm neut}} \right]^{1/3} \,\mbox{: outer radius of the shell}
\end{equation}
The outer
radius $R_{\rm out}$ of the neutral part of the shell -- with mass
$M_{\rm neut} = M_{\rm S}-M_{\rm ion}$ -- is calculated assuming a
uniform average density $\rho_{\rm neut}$, 
which is derived from the equation of state of perfect gas,
assuming a temperature of 100 K. In this case the isothermal speed of sound
is typically of 1 km s$^{-1}$.

In general, we have $R_{\rm S}^{*} \le R_{\rm ion} \le R_{\rm out}$,
the first and second equalities corresponding to the cases of complete
shell neutrality ($M_{\rm neut} = M_{\rm S}$) and ionisation 
$(M_{\rm neut} = 0$), respectively.

Moreover, it follows that 
the expansion of the ionised part of the shell
produces a deceleration of the actual contact discontinuity 
that will expand with a velocity 
	\beq
	V_{\rm S}^* = V_{\rm S} + \dot{\Delta R_{\rm b}} \,\,\,\,
	\,\,\,\,\,\,\,\, (\mbox{with}\,\,\,\dot{\Delta R_{\rm b}} < 0) \, ,
	\eeq
and an acceleration  of the ionisation  front
that will move outward with a velocity
	\beq
	V_{\rm ion} = V_{\rm S} + \dot{\Delta R_{\rm a}}  \,\,\,\,
	\,\,\,\,\,\,\,\, (\mbox{with}\,\,\,\dot{\Delta R_{\rm a}} > 0)
	\eeq
%

%

It should to be noticed that 
such a first-order perturbation approach 
is strictly correct only
as long as $\Delta R_{\rm b} \ll R_{\rm S}$, i.e.
during the early stages of the expansion when a small fraction of
the shell is ionised.  
Moreover, with the above prescriptions 
we implicitly suppose that there is a
pressure equilibrium and homogeneity in the ionised part of
the shell. This assumption is acceptable considering 
that a rarefaction wave,
generated by the initial inward motion of the contact discontinuity, is
expected to rapidly homogenise the gas (Breitschwerdt \& Kahn 1990).
It is also clear that the adopted treatment is 
a very simplified approximation of the real phenomenon. 
Nevertheless, comparing our results with both 
the predictions of hydrodynamical models 
and observed data,  we generally find a good agreement
(see Figs.~\ref{fig_sthick} and \ref{fig_thick}).

An example of the application of the 
shell-thickening approximation is shown in Fig.~\ref{fig_sthick},
that refers to a model nebula surrounding a central star
with mass of $\sim 0.6 \, M_{\odot}$, and initial solar metallicity.

The relative thickness of the ionised shell 
	\beq
\delta R = 
	\frac{R_{\rm ion}-R_{\rm S}^*}{R_{\rm ion}}
	\label{eq_dr}
	\eeq
is shown as a function of time (solid line, bottom-left panel), and 
compared to that expected according to the thin-shell approximation
(dashed line).
The latter curve is derived by integrating 
Eqs.~(\ref{eqr}) -- (\ref{eqp}), and assuming
that the gas density in the ionised shocked shell is 
a factor $k = 5$ times larger than the density of the unperturbed 
AGB wind, i.e. 
$\rho_{\rm ion} = k \dot M_{\rm AGB} / (4 \pi R_{\rm S}^2 V_{\rm S})$,
as suggested by VK85.\footnote{$k=5$ is indicated by VK85 as a reasonable 
compromise between the high density contrast generated by
 a isothermal shock and the density decrease in the ionised nebula}
In this case the thickness slowly increases with time,
being mostly below 0.1.

On the contrary, when the pressure contribution 
due to ionisation is taken into account,
the shocked shell progressively thickens as the ionisation front
breaks into it, reaching a maximum 
of $\delta R \sim 0.7$ when the whole shell is ionised.
Actually, at this stage, the ionisation front outpasses the outer rim
of the shocked shell and embraces the undisturbed low-density 
AGB wind completely (see top-left panel of Fig.~\ref{fig_sthick}).

After the maximum of $\delta R$,
its subsequent decline and further increase
can be understood considering the competition between the rates of
i) the mechanical energy input from the fast wind -- which tends to compress 
the shell from the back (term $\dot m v^2/2$) --  and ii) the work done
by the pressure to expand the shell.
The latter is expressed by the 
term $- \dot E_{\rm W} = 4 \pi P R_{\rm S}^2 V_{\rm S}$  for the 
unperturbed case, and the equivalent term 
$- \dot E_{\rm W}^{*} = 4 \pi (P+\Delta P_{\rm ion})
 (R_{\rm S} + \Delta R_{\rm b})^2  (V_{\rm S} + \Delta \dot R_{\rm b})$
when the first-order corrections are applied.
As shown in Fig.~\ref{fig_sthick} (top-right panel), 
as long as the first term dominates, the shell thickness 
decreases; at later times it starts to increase again, 
as the kinetic energy input of the fast-wind drops and  
the adiabatic expansion prevails. 

Similar considerations can be applied  to explain 
the temporal behaviour of the velocity
at the inner and outer rims of the ionised nebula (bottom-right panel).
The increase of the pressure inside of the ionised shell determines 
the initial strong acceleration of the ionisation front and 
the deceleration of the contact discontinuity (which may even attain
negative values corresponding to a short-lived inward motion).
The subsequent acceleration of both rims 
(the shell is now fully ionised) is driven by the increasing 
pressure of the hot bubble. Finally, when the strength of the 
fast wind weakens, the contact discontinuity and the outer rim 
of the nebula are decelerated and accelerated, respectively, 
while the ionised gas 
tends to expand both inward and outward.
We can note that the velocities attained at the same time 
by the shell rims can be quite 
different, and cover the wide range of observed velocities in PNe,
typically from few km s$^{-1}$ to 30 -- 40 km s$^{-1}$.     

The reliability of these predictions can be checked by comparison 
with the results of hydrodynamical calculations from
Marten \& Sch\"onberner (1991). 
As far as the shell thickening is concerned, 
it is encouraging to note that the temporal behaviour of $\delta R$
shown in Fig.~\ref{fig_sthick}
closely resembles the predictions of their numerical simulations 
(see figure 10 in Marten \&  Sch\"onberner).
The evolution of the rim velocities -- i.e.  
the acceleration of the ionisation front and deceleration of the 
contact discontinuity during the early stages of the ionisation phase
-- is also recovered by our analytical model.

\begin{figure*}
\begin{minipage}{0.69\textwidth}
\resizebox{\hsize}{!}{\includegraphics{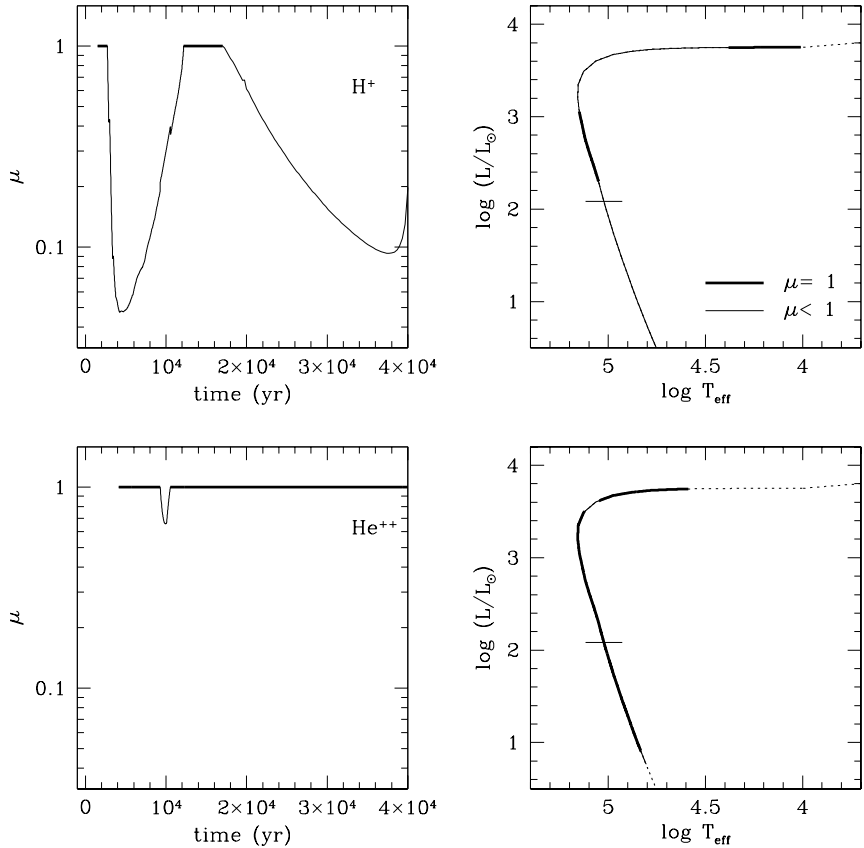}}
\end{minipage}
\hfill
\begin{minipage}{0.30\textwidth}
\caption{Expected optically thick and thin configurations during the
evolution of a PN ionised by a central star with a mass of $\sim 0.6
M_{\odot}$ and initial solar composition  
(model 4 in Table~\protect{\ref{tab_mod}}). 
{\bf Left panels:} absorbing factor $\mu$ 
(given by Eq.~\protect{\ref{eq_mu}})
as a function of time for both H$^{+}$ and He$^{++}$ ions (top
and bottom panels, respectively).
Time is set to zero at the end of the AGB evolution.
{\bf Right panels:} Predicted optical appearance of the nebula 
as the central star evolves along the post-AGB track on the H-R
diagram. The adopted notation is as follows: 
dotted line as long as the photoionisation rate $Q < 10^{40}$
s$^{-1}$; solid lines for optical thick ($\mu = 1$) or thin 
($\mu < 1$) configurations as indicated. 
The horizontal mark on the tracks corresponds to 
the evolutionary age of $4\times10^4$ yr, which corresponds
to the limit of the interval shown in the left plots}
\label{fig_mu}
\end{minipage}
\end{figure*}

\subsection{Optical properties of the nebulae}

With the aid of the prescriptions given 
in Sect.~\ref{ssec_ionmas}, we are able to determine 
not only the ionised masses
for H$^{+}$, He$^{+}$, and He$^{++}$ at each time step during the 
post-AGB evolution, but we can also naturally discriminate between 
{\sl optically thick} and  {\sl optically thin} configurations,
depending on whether the continuum photons (relative to a given ion) are 
partially or totally absorbed by radiative recombination transitions 
in the nebula. To this aim, we calculate at each time during 
the evolution the {\sl absorbing factor} (as defined by 
M\'endez et al. 1993)
\beq
\label{eq_mu}
	\mu = \frac{N_{\nu}^{\rm abs}}{N_{\nu}^{\rm em}} 
\eeq
giving  the fraction of the continuum emitted photons that is actually
absorbed by the ion species under consideration.
In practice, it can be expressed as the ratio 
between the ionisation rate $Q$ and the recombination rate within
the Str\"omgren radius, $R_{\rm ion}$ 
(see Eqs.~\ref{eq_qh} -- \ref{eq_qhep}).
It follows that $\mu = 1$  as long as
$R_{\rm ion} \le R_{\rm neb}$, which corresponds to the 
optically thick case. On the other hand, if
no radial coordinate, up to  $R_{\rm neb}$, 
can be found where the equilibrium condition is satisfied    
-- so that $R_{\rm ion}$ cannot
be formally defined --, we have  $\mu < 1$, which corresponds to the 
optically thin case.
Figure \ref{fig_mu} illustrates how the nebular optical
properties in both H and He$^{+}$ continua  are predicted to vary
as a function of time as the central star evolve on its post-AGB
track. 

As far as the H$^{+}$ continuum is concerned, we can see
that when the central star becomes hot enough to 
ionise the H atoms, the nebula is expected to be initially optically
thick. The transition to the optically-thin 
configuration occurs as the effective temperature keeps increasing 
along the horizontal part of the track and the number of the emitted 
ionising photons is growing. At this stage the whole nebula
is ionised. 
In proximity to the knee of the track 
(the point of maximum temperature), $\mu$ attains its minimum value,
implying that a large number of photons are escaping from the ionised
nebula. Subsequently, the luminosity starts to drop, the 
photoionisation rate decreases (hence $\mu$ increases), and the 
nebula becomes optically thick again.  

Later on,  a second conversion to the optically-thin configuration 
occurs, being essentially the result of the dynamical expansion of the nebula. 
In fact, when the gas density gets low enough, recombination 
in the whole nebula becomes less efficient (in terms of rate of events)
than photoionisation.

The optical properties in 
the  He$^{++}$ show notable differences compared to the
case of H$^{+}$. First, helium starts to be doubly ionised 
at quite high effective temperatures (typically $T_{\rm eff} \sim 60\,000$ K).
Second, the optically-thin configuration is 
very short-lived, and restricted 
to the knee of the track. This reflects the fact that the photoionisation
rate for He$^{+}$ is systematically lower than that for H 
(see Fig.~\ref{fig_ionf}), and presents a much steeper decrease 
with effective temperature and luminosity. 

Finally, it is worth remarking that in our model ionisation and optical
properties of PNe are predicted according to a coherent coupling 
between the evolution of the central star and the dynamics of the
nebula. 
This represents an important improvement upon past synthetic 
approaches, that in many cases lack any dynamical description of 
the nebulae (e.g. Stanghellini \& Renzini 2000) and/or adopt  
$\mu$ as free parameter (e.g. M\'endez \& Soffner 1997).

\subsection{Intensities of few relevant optical emission lines}
\label{ssec_lines}
We aim at predicting  
the nebular luminosities at the wavelengths of 
a few important lines typically observed in PNe, namely:
the recombination lines  H$\beta\,\lambda4861$ and 
He{\sc \,ii}$\,\lambda4686$, and the nebular forbidden line
[O{\sc \,iii}]$\,\lambda 5007$.  

As a first approach, the intensities of these lines 
are calculated by standard procedures (as described below), 
under simple but reasonable
assumptions for the physical conditions 
of the emitting regions.
Specifically, we assume that the temperature $T_{\rm e}$ 
and density $N_{\rm e}$
of free electrons are constant.
In particular, $N_{\rm e}$ is predicted according to the ionisation
structure described by Eqs.~(\ref{eq_ion1}) -- (\ref{eq_ion3}), whereas
$T_{\rm e}$ is considered a free input parameter, which is chosen
in the typical range indicated by observations, that is 
$8\,000\,{\rm K}\,  \la  T_{\rm e} \, \la 15\,000 \, {\rm K}$.

These assumptions may be considered completely satisfactory
as far as recombinations lines are concerned, as they depend mostly
on atomic parameters and only weakly on $T_{\rm e}$. 
The case of forbidden lines is
more delicate, due to their high sensitiveness 
the electron temperature which
cannot be predicted but has to be assumed by the model.
It is clear that an improved description of the 
line emission from PNe can be attained only 
with the aid of a photoionisation code.  
However, our choice should represent a reasonable 
first step, and should be adequate to the purpose of exploring 
the sensitiveness of PNe properties to various possible physical 
conditions.

\subsubsection{The recombination lines: H$\beta$ and 
He{\sc \,ii}$\,\lambda 4686$}
The evaluation of the total luminosity $L_{\lambda}$ emitted by the ionised
nebula in a particular recombination transition 
(of wavelength $\lambda$) requires the calculation of
	\beq
	L_{\lambda} = 4 \pi \, \int_{R^*}^{R_{\rm ion}} J_{\lambda}(r)
	r^2 {\rm d} r
	\eeq
where  the integral is performed over the Str\"omgren volume of the 
ion under consideration, with 
	\beq
	J_{\lambda} = \alpha(n n')\, N_{\rm e} \, N_{\rm ion}\, h\nu
	\eeq
being the volume emission (erg cm$^{-3}$ s$^{-1}$), and $\alpha(n n')$
denoting the recombination coefficient of the transition
from state $n$ to state $n'$. 

This latter parameter is calculated adopting the fitting relations 
presented by P\'equignot et al. (1991), as a function of $T_{\rm e}$
and ionic charge, for both the recombination
H$\beta$ line of H at $\lambda = 4861$ \AA~~[$\alpha_{{\rm H}^+}(4,2)$], 
and the recombination line of He{\sc \,ii} at $\lambda = 4686$ \AA~~
[$\alpha_{{\rm He}^{++}}(4,3)$].


All other quantities in the above equations are known once the
ionisation structure of the nebula is determined following
the prescriptions described in Sect.~\ref{ssec_ionmas}. 

\subsubsection{The forbidden line: [O{\sc \,iii}]$\,\lambda 5007$}
The intensity of this nebular line 
-- corresponding to the transition from the collisionally-excited
metastable level $j=\, ^{1}{\rm D}_2$ to the level $i=\, ^{3}{\rm P}_2$ -- 
is calculated from
	\beq
	\label{eq_oii}
I_{ji} = A_{j i}\,  N_j\,  \frac{Y({\rm O}^{++})}{Y({\rm O})}\, 
	\frac{Y({\rm O})}{Y({\rm H})}\, h \nu_{ji}\,
	\frac{n}{1 + Y({\rm He})/Y({\rm H})}
	\eeq  %
(expressed in erg~cm$^{-3}$~s$^{-1}$), 
where $A_{j i} = 0.021$~s$^{-1}$ is the radiative transition probability; 
$N_{j}(T_{\rm e}, N_{\rm e})$ is the relative population of level
$j$; $Y({\rm O}^{++})/Y({\rm O})$ denotes the ionic abundance
of ${\rm O}^{++}$ relative to the total oxygen abundance 
(in all excitation and ionisation states);  
$Y({\rm O})/Y({\rm H})$ refers to  
abundance of oxygen (in all states) with
respect to hydrogen; $Y({\rm H})$ and $Y({\rm He})$ are the chemical
abundances (by number, in mole~gr$^{-1}$) of hydrogen and helium;
and $n = N({\rm H}) + N({\rm He})$ corresponds to the number density
(in cm$^{-3}$) of H and He.

\begin{table*}
\caption{Specification of the basic parameters adopted to calculate
PN models}
\label{tab_mod}
\begin{tabular}{llllllll}
\hline\noalign{\smallskip}
$N_{\rm mod}$  & $M_{\rm 1TP}$ &
$M_{\rm CSPN}$ & H-/He- & $V_{\rm AGB}^{\rm
max}$  & $t_{\rm tr}$ & $T_{\rm e}$ & notes \\
  & $(M_{\odot})$ &
$(M_{\odot})$ & burner & $({\rm km}\,{\rm s}^{-1})$ & $(10^3\,{\rm
yr})$ & $(10^3\, {\rm K})$ &    \\
\noalign{\smallskip}\hline\noalign{\smallskip}
1  & 1.7 & 0.5989 & H  & 15     & 1.5 & 10.0 &  C \\
2  & 1.7 & 0.5989 & H  & $f(Z)$ & 1.5 & 10.0 &  C \\
3  & 1.7 & 0.5989 & H  & $f(L)^*$ 
& 1.5 & 10.0 & C \\
4  & 1.7 & 0.5989 & H  & $f(L)$ & 1.5 & 10.0 &  A, B, C, D, E\\
5  & 1.7 & 0.5989 & H  & 15     & 0.5 & 10.0 &  C, D\\
6  & 1.7 & 0.5989 & H  & $f(L)$ & 1.5 & 10.0 & thin-shell approx. \\
7  & 1.7 & 0.5989 & H  & $f(L)$ & 1.5 & 10.0 & blackbody \\
8  & 1.7 & 0.5989 & H  & $f(L)$ & 8.0 & 10.0 &  D \\
9  & 1.7 & 0.5989 & He & $f(L)$ & 1.5 & 10.0 &  B, D\\
10 & 1.7 & 0.5989 & He & $f(L)$ & 5.0 & 10.0 &  D \\
11 & 1.7 & 0.5989 & H  & $f(L)$ & 1.5 &  8.0 &  E \\
12 & 1.7 & 0.5989 & H  & $f(L)$ & 1.5 & 12.0 &  E \\
13 & 1.7 & 0.5989 & H  & $f(L)$ & 1.5 & 13.0 &  E \\
14 & 1.0 & 0.5577 & H  & $f(L)$ & 10.0 & 10.0 & C  \\
15 & 1.0 & 0.5577 & H  & $f(Z)$ & 10.0 & 10.0 & C  \\
16 & 1.2 & 0.5706 & H  & $f(L)$ & 1.5  & 10.0 & A, D \\
17 & 1.2 & 0.5706 & H  & $f(L)$ & 8.0  & 10.0 & D  \\
18 & 1.5 & 0.5868 & H  & $f(L)$ & 1.5  & 10.0 & A \\
19 & 1.5 & 0.5868 & He & $f(L)$ & 10.0  & 10.0 &  \\
20 & 1.8 & 0.6120 & H  & $f(L)$ & 1.5  & 10.0 & A \\
21 & 2.0 & 0.6296 & H  & $f(L)$ & 1.5  & 10.0 & A, B ,D \\
22 & 2.0 & 0.6296 & H  & $f(L)$ & 0.5  & 10.0 & D \\
23 & 2.0 & 0.6296 & He & $f(L)$ & 1.5  & 10.0 & B \\
24 & 2.2 & 0.6529 & H & $f(L)$ & 1.5  & 10.0 &  A, D \\
25 & 2.2 & 0.6529 & H & $f(L)$ & 0.5  & 10.0 &  D \\
26 & 2.5 & 0.6865 & H  & $f(L)$ & 1.5  & 10.0 & A, D \\
27 & 2.5 & 0.6865 & H  & $f(L)$ & 0.5  & 10.0 &  D \\
28 & 2.5 & 0.6865 & H  & $f(L)$ & 0.3  & 10.0 &  D \\
29 & 3.0 & 0.7867 & H  & $f(L)$ & 0.5  & 10.0 &  \\
\noalign{\smallskip}\hline
\end{tabular}

\bigskip

Sequences of models with the same set of
model parameters, but varying\\
A: $M_{\rm CSPN}$ (4-16-18-20-21-24-26), (9-16); \\ 
B: H-/He-burning post-AGB tracks (4-9), (8-16); \\
C: $V_{\rm AGB}^{\rm max}$ (1-2-3-4-5), (14,15); \\
D: $T_{\rm tr}$ (4-8),(9-10), (21-22), (24-25), (26-27-28); \\
E: $T_{\rm e}$ (4,11,12,13).\\

$^*$ According to the original results of Kr\"uger et al. (1994)
(i.e. $f_{\rm v}=1$; see Sect.~\protect{\ref{sssec_vagbm}}); in all other
models $f_{\rm v} = 0.588$ is adopted.   
\end{table*}

The most critical quantity here
is $N_{j}(T_{\rm e}, N_{\rm e})$, as it 
depends on both the electron temperature and density.
It is determined using the detailed calculations 
of atomic transitions performed by Kafatos \& Lynch (1980),  
interpolating in $T_{\rm e}$ and $N_{\rm e}$ the data of their Table 3E.
All other quantities (i.e. elemental abundances and densities) 
are predicted by our model.

Equation~(\ref{eq_oii}) must be integrated over volume to obtain 
the emitted luminosity in the [O{\sc \,iii}]~$\lambda 5007$ line.
We assume that both O$^{++}$ and  He$^{+}$ ions share 
the same Str\"omgren volumes, 
based on the fact the ionisation potentials of O$^{++}$
(54.9~eV) and  He$^+$ (54.4~eV) are almost identical. 
In other words, we assume that the emission of the 
[O{\sc \,iii}]~$\lambda 5007$ line can only originate from 
regions where He is single-ionised, disappearing as soon
as He becomes double-ionised.
Moreover, we consider that within that region 
$Y({\rm O}^{++}) =Y({\rm O})$, which is also a good approximation
(Osterbrock 1974). 


\subsection{Summary of model parameters}
\label{sec_par}
Before analysing the expected 
dynamical and ionisation properties of PNe, it is worth summarising  
the parameters/prescriptions of the model
that are can be varied in our calculations, namely:    
\begin{itemize}
\item initial metallicity of the stellar progenitor, $Z_{\rm i}$;
\item stellar mass 
at the onset of the TP-AGB phase, $M_{\rm 1TP}$;
\item mass of the central star of the PN model, $M_{\rm CSPN}$; 
\item H- or He-burning post-AGB track; 
 \item the transition time, $t_{\rm tr}$; 
\item expansion velocity of the stellar wind during the last stages
of the AGB evolution, $V_{\rm AGB}^{\rm max}$; 
\item electron temperature of the ionised shell, $T_{\rm e}$.
\end{itemize}

It is worth remarking that our default set of prescriptions
include i) the ionisation fluxes from model atmospheres, and ii)
the thick-shell correction scheme. 
For purpose of comparison, we may easily relax these assumptions 
adopting the blackbody energy distribution and the thin-shell approximation.
This is the case for few examples discussed in the text.

Calculated PNe models are listed in Table~\ref{tab_mod}, so that 
each reference number N$_{\rm mod}$ corresponds to a particular
set of parameters.
It is clear that some of these quantities are intrinsic properties
of the models (e.g. $Z_{\rm i}$, $M_{\rm CSPN}$), whereas others are
genuine  {\sl free} parameters (e.g. $t_{\rm tr}$, 
$V_{\rm AGB}^{\rm max}$, $T_{\rm e}$).
The aim is to define suitable ranges for these parameters, 
on the basis of observations of Galactic PNe. 
For the other model ingredients, not explicitly mentioned here, 
we refer to the prescriptions already discussed in Sect.~\ref{sec_pnmod}.

\section{Overview of model predictions}
\label{sec_overview}
\begin{figure*}
\centering
\resizebox{0.8\hsize}{!}{\includegraphics{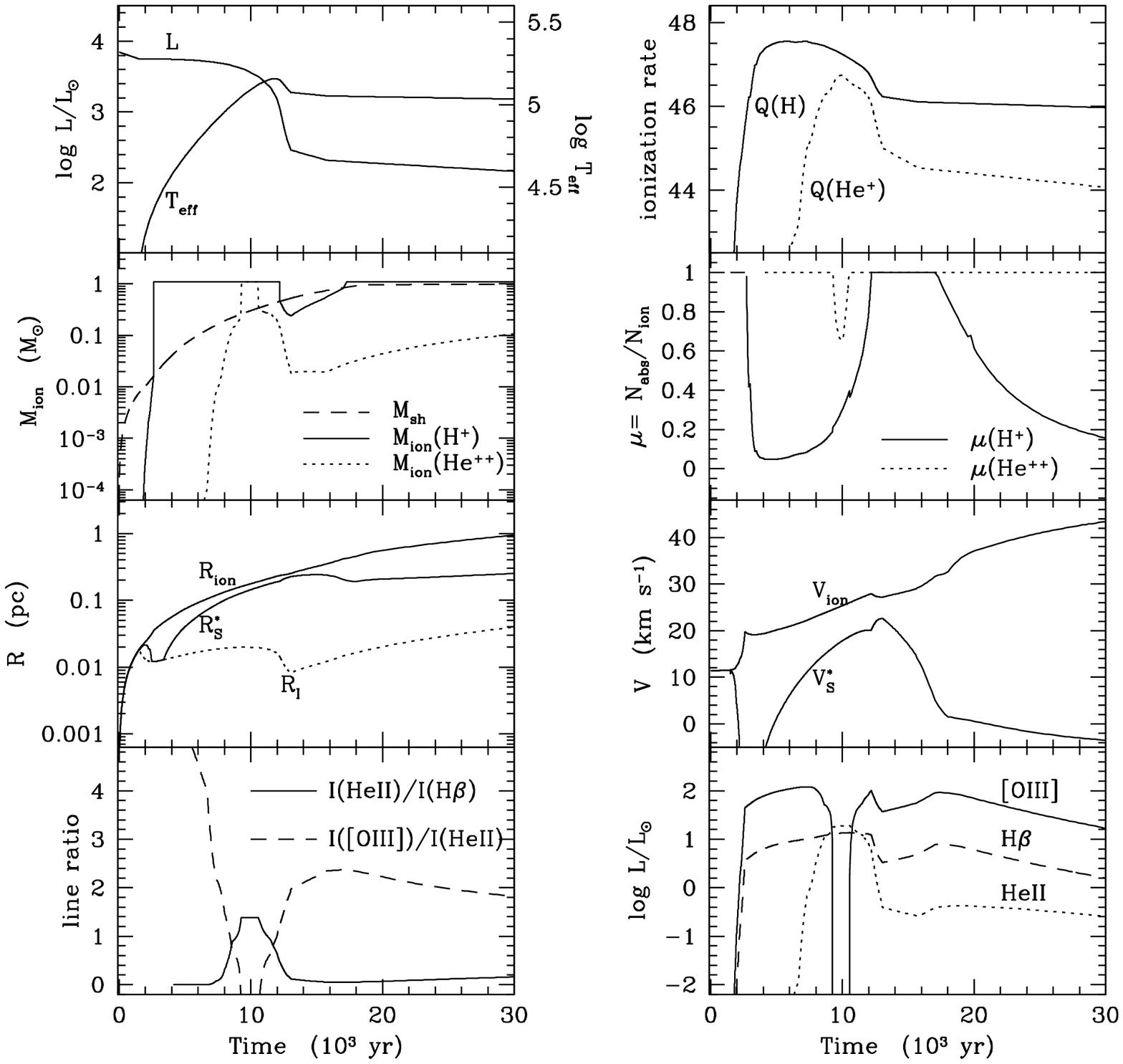}}
\caption{Predicted evolution of basic parameters corresponding to a PN
model ionised by a H-burner central star with mass of $\sim 0.6\,
M_{\odot}$ (model 4). See Sect.~\protect\ref{sec_overview} for a
description  }
\label{fig_pn17h} 
\end{figure*}

\subsection{Summary plots}

The several panels in Fig.~\ref{fig_pn17h} illustrate the time
evolution -- during a period of 30\,000~yr -- of a PN model 
corresponding to a H-burning central star with a mass 
of about 0.6~$M_{\odot}$. The selected model 
(number 4 in Table~\ref{tab_mod}) 
may be considered typical of Galactic PNe. 
The run of several physical quantities is
shown simultaneously, then offering a useful guide to our further
discussion. In short, going from top to bottom panels, we present:
	\begin{enumerate}
	\item
the evolution of luminosity and temperature for the central star
(left; Sect.~\ref{sec_tracks}), and the corresponding ionisation rates 
(right; Sect.~\ref{sec_photrates}); 
	\item 
the masses of the shocked shell and ionised regions (left), and the 
corresponding $\mu$ factors (right; Sect.~\ref{ssec_ionmas});
	\item
the radii (left) and velocities (right) of  
the shell boundaries (including the radial location 
of the inner shock) (Sect.~\ref{ssec_thick}); 
	\item 
the line ratios (left) and total luminosities (right) in the main
emission lines (Sect.~\ref{ssec_lines}).
	\end{enumerate}

\begin{figure*} 
\centering
\resizebox{0.8\hsize}{!}{\includegraphics{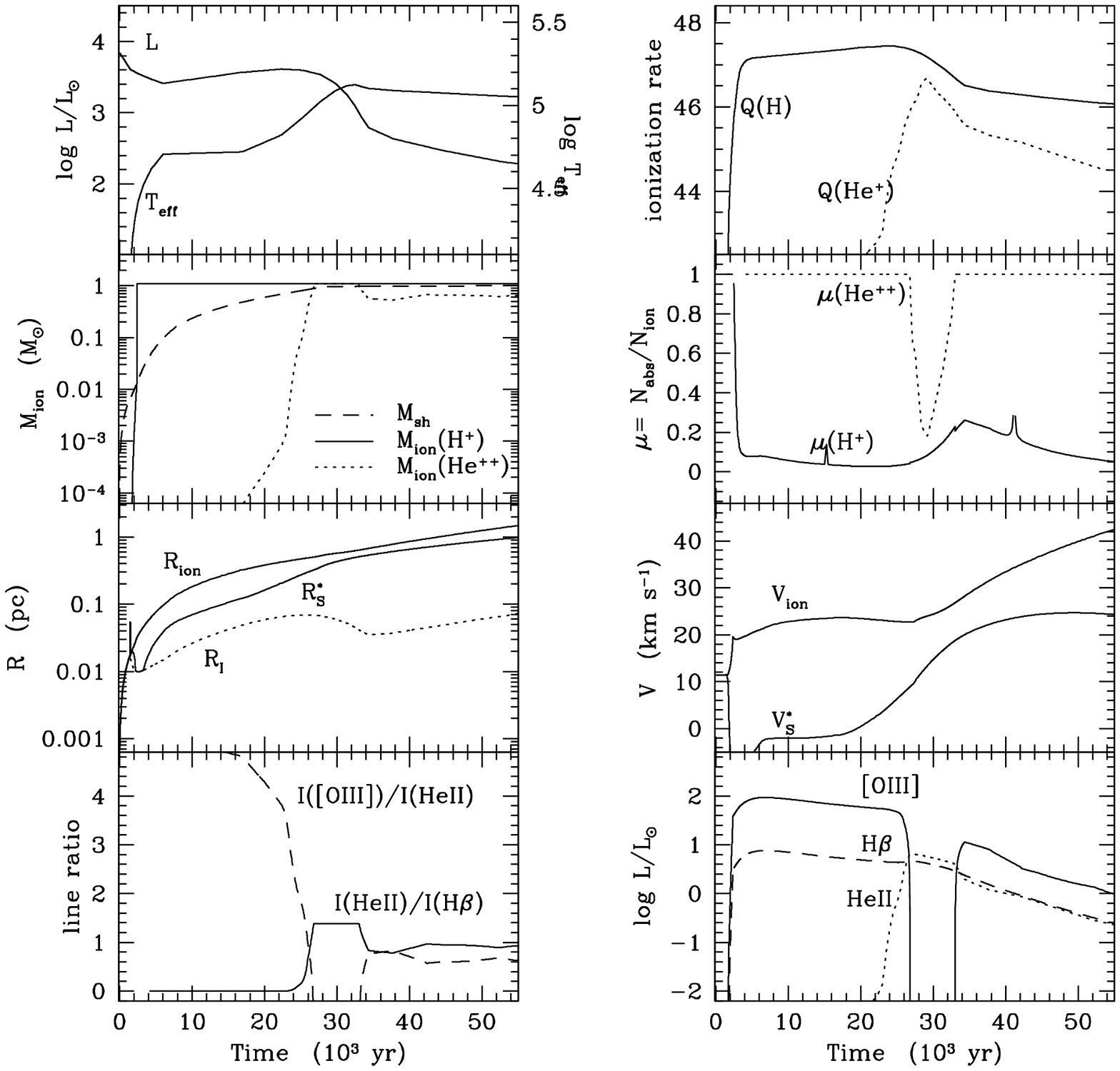}}
\caption{The same as in Fig.~\protect{\ref{fig_pn17h}},
but for a He-burner central star with mass of $\sim 0.6\,
M_{\odot}$ (model 9)  
}
\label{fig_pn17he} 
\end{figure*}

Similar plots are available for all our models, and allow a better
interpretation of the different evolutionary features that may be
present. A very interesting comparison is provided, for instance,
between the H-burning track of Fig.~\ref{fig_pn17h}, and a model
computed with the same set of parameters, but using a He-burning track (number
9 in Table~\ref{tab_mod}).  The latter is presented in
Fig.~\ref{fig_pn17he}, for a total period of 55\,000~yr.  As can be
readily noticed, the He-burning track is characterised by a smaller
luminosity and slower evolution than the H-burning one, and this 
determines a substantially different evolution of the nebular
properties. For instance, the nebula shown in Fig.~\ref{fig_pn17he}
is nearly always optically thin to H-ionising photons ($\mu({\rm
H}^{+})<1$), and presents line ratios [O{\sc \,iii}]$\,\lambda
5007$/He{\sc \,ii}$\,\lambda4686$ and He{\sc
\,ii}$\,\lambda4686$/H$\beta$ that at late times are
very different from those predicted for  the PN model of Fig.~\ref{fig_pn17h}.
These differences are mainly due to the fact that, 
in comparison to the H-burning central star, 
the maximum flux of ionising photons emitted 
by the He-burning star is attained
later (because of the longer evolutionary 
time-scales), when the nebular material has expanded to larger radii.

\subsection{Detection probabilities of different PNe}

The relative probability of observing PNe of different progenitors
(i.e. stars of different masses, ages, and metallicities) constitutes
a basic aspect in the interpretation of PNe data.  As briefly
mentioned in Sect.~\ref{intro}, the highest probabilities of observing
the PN stage should be reached when we have a suitable combination
between the evolutionary time-scale of the central star, and the
dynamical time-scales of the expanding nebulae: A fine tuning between
the two can increase either the lifetime or the surface brightness of
a PN, both factors enhancing the detection probability. 

\begin{figure*} 
\begin{minipage}{0.48\textwidth}
\resizebox{\hsize}{!}{\includegraphics{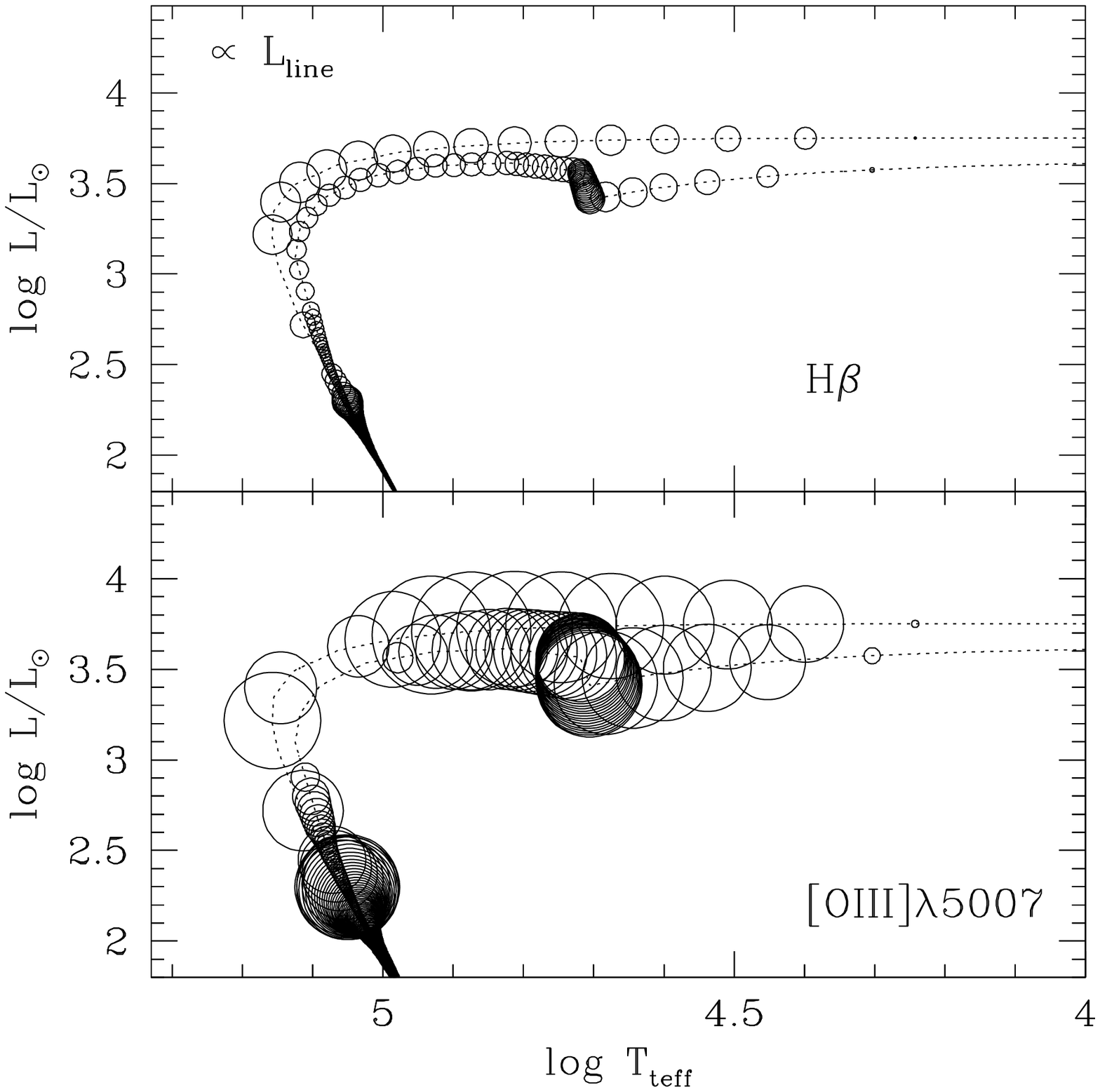}}
\end{minipage}
\hfil
\begin{minipage}{0.48\textwidth}
\resizebox{\hsize}{!}{\includegraphics{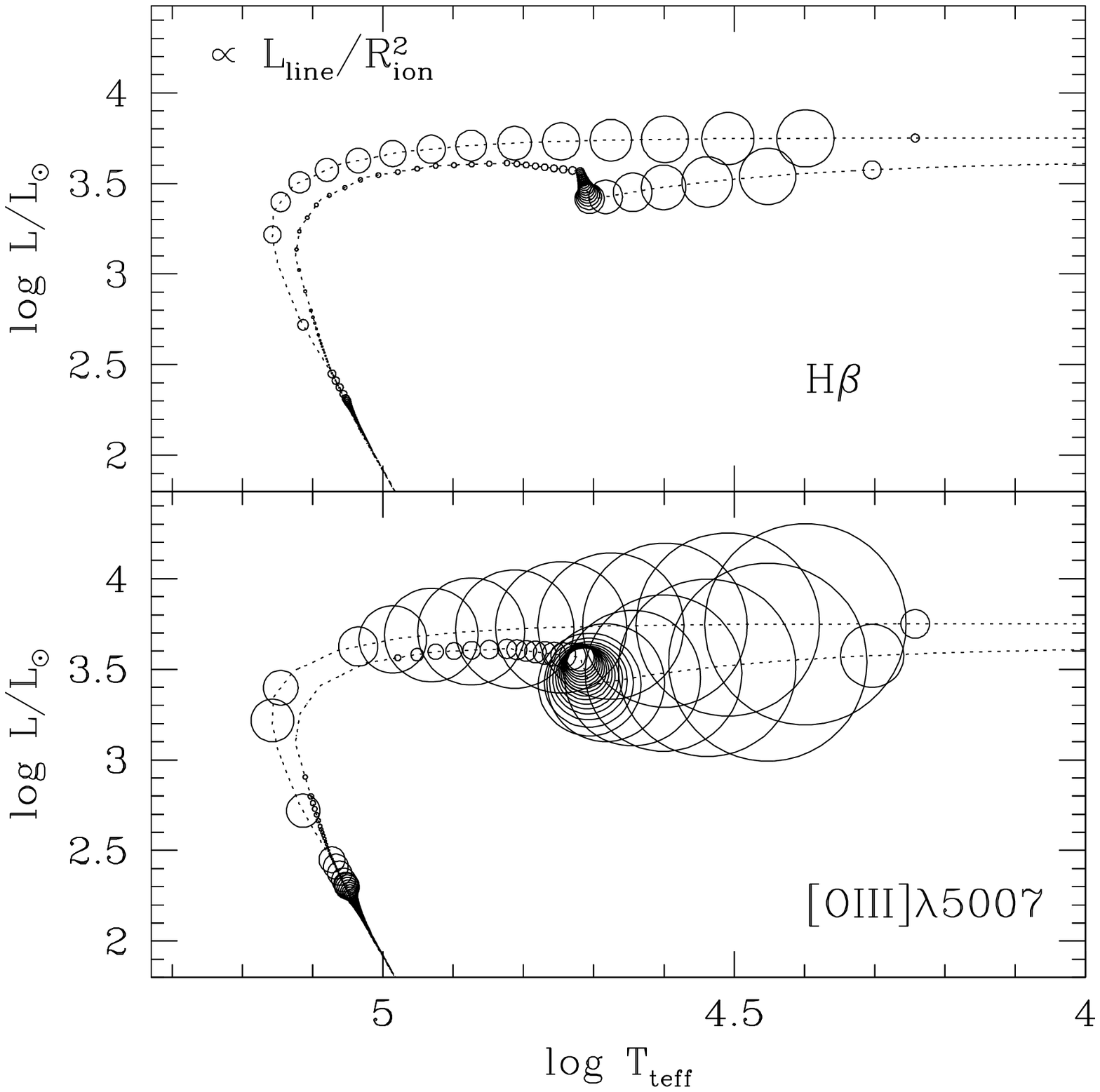}}
\end{minipage}
\caption{Representation of the relative probabilities to detect a PN
observed either in H$\beta$ (top panel) or in the [O{\sc\,
iii}]$\,\lambda5007$ line (bottom panel), as its central star evolves
along its post-AGB track, assumed either H-burning (model 4) or
He-burning (model 9, that corresponds  to the least luminous track 
in each panel).
Circles are drawn at equally-spaced time intervals of 700~yr,
their area being proportional to the current values of the luminosity 
(left panels), and surface
brightness (right panels) in each emission line}
\label{fig_pnprob} 
\end{figure*}

Actually, the problem of determining a ``detection probability'' is
much more complex. For instance, in the case of distant (point-like)
nebulae, the detection probability should correlate with the
absolute luminosity, rather than with the surface brightness. Also the
choice of the detection method is crucial in determining which PNe will be
better sampled by a given survey. Since our models predict a lot of
different properties of PNe, we are also able to face this question in
a more generic way: we can construct synthetic samples of PNe
containing all their basic observable properties, which are then be
filtered adopting some given observational criteria. Developing such a
tool, however, is beyond the scope of this paper.

Anyway, it is worth getting a first insight into the problem. To this
aim, in Fig.~\ref{fig_pnprob} we plot the evolution of two PNe models
in the HR diagram, corresponding to H- and He-burning tracks (models 4
and 9, respectively).  Along the tracks, we mark the CSPN position at
equally spaced intervals of age (700~yr) by means of circles whose
area is proportional either to the emitted line luminosity $L_\lambda$
(left panels) or to the PN average surface brightness $S_\lambda$
(right panels) in a given emission line (either H$\beta$ or [O{\sc
\,iii}]$\,\lambda5007$).  $S_\lambda$ is simply assumed to be
proportional to the quantity $L_\lambda/R_{\rm ion}^2$, where $R_{\rm
ion}$ is the total nebular radius.

The figure gives a clear indication of the regions of the HR diagram
where these PNe are more likely to be found, as well as of their
relative luminosities and surface brightnesses. 
In the case of the H-burning track, the
probability of detection in the H$\beta$ line should be more or less
evenly distributed in effective temperature, as long as
$\log\Teff\ga4.2$ and the CSPN is crossing the HR diagram
horizontally: In this part of its evolution, the nebula evolves at
slightly increasing $L_{\rm H\beta}$, and slightly decreasing 
$S_{\rm H\beta}$.  Also, there is a non-negligible probability of
finding these PNe at later ages, when their central
stars reach the initial part of the white dwarf cooling sequence (at
$\log L/L_\odot\simeq2.3$, $\log\Teff\simeq5.05$) and the evolutionary
speed slow down. However, in this part of the HR diagram, $S_{\rm H\beta}$
is extremely low.  

The same H-burning nebula should be much brighter in the [O{\sc
\,iii}]$\,\lambda5007$ line. Remarkably, during the horizontal crossing of
the HR diagram, the detection of [O{\sc \,iii}] would be favoured in a
more limited temperature interval, i.e. $5.0\ga\log\Teff\ga4.2$; at
higher temperatures, the [O{\sc \,iii}] line temporarily disappears. 
This effect is caused by the photoionisation of O$^{++}$ into
O$^{+++}$, 
and is explained in more detail in Sect.~\ref{sec_lineratios}.

Figure~\ref{fig_pnprob} also illustrates the striking differences
between the H- and He-burning cases. The PN with the He-burning track
have line luminosities that are just slightly lower than its H-burning
counterpart. However, except for the very initial stages, the He-burning
model presents much lower surface brightnesses. This is caused by
the longer evolutionary time-scales of the He-burning central stars.
Moreover, the He-burning CSPN makes a long pause in its evolution amid
the horizontal crossing of the HR diagram, at $\Teff\simeq4.7$, period
in which the nebula should have a very high probability of being
detected. During this pause, the line luminosity is almost constant,
but the surface brightness decreases a lot, due to nebular
expansion. Thereafter, the CSPN resumes its evolution at a slower pace
than in the H-burning case. In summary, PN with He-burning tracks 
should present a very odd distribution in the HR diagram: most of them
should concentrate in the $\log\Teff\ga4.7$ region, with a small
fraction of high-surface brightness objects appearing at
$4.7\ga\log\Teff\ga4.3$. This trend should be present in both H$\beta$
and [O{\sc \,iii}]$\,\lambda5007$ lines. However, similarly to the
H-burning case, the [O{\sc \,iii}] line disappears in the upper-left
section of the HR diagram, when $\Teff\ga4.9$ and $\log L/L_{\odot} \ga 3$.

\section{Comparison with observations}
\label{sec_observ}

The next sections are dedicated to present the results of PNe model
calculations in comparison to the observed data, 
which are used to both calibrate the free parameters and, 
more generally, test the theoretical
prescriptions. As to the calibration of the parameters,   
we restrict the comparison here to a sample of  Galactic PNe,
so as to minimise possible population effects 
(due e.g. to different mean metallicity) that may be present among 
PNe samples of nearby galaxies.  

\subsection{The ionised mass -- nebular radius relationship}
\label{sssec_masrad}

\begin{figure}
\resizebox{\hsize}{!}{\includegraphics{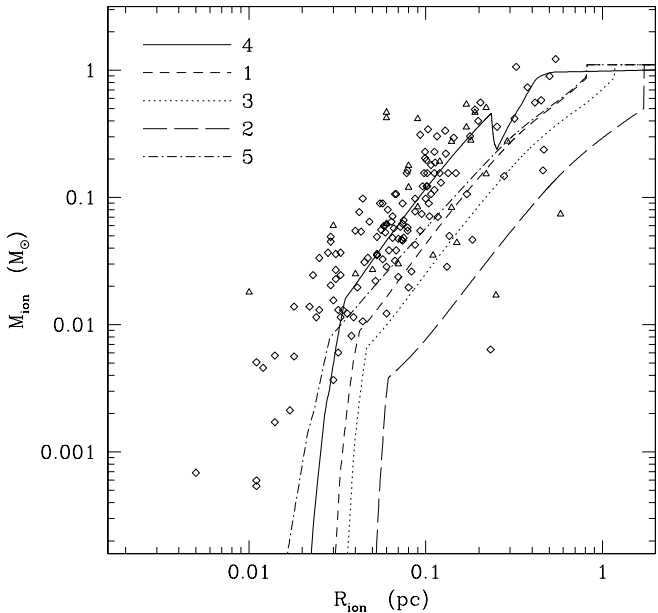}}
\caption{Ionised mass -- nebular radius relationship.
Observed data for Galactic PNe are taken from the compilation
by Zhang (1995; squares), and the sample presented 
by Boffi \& Stanghellini (1994; triangles). Model predictions
are displayed for various prescriptions for $V_{\rm AGB}^{\rm max}$ 
and $t_{\rm tr}$ (see Table~\protect{\ref{tab_mod}})}
\label{fig_masradcal}
\end{figure}

PNe display a  positive
correlation between the ionised mass and the  nebular
size, which is shown in Fig.~\ref{fig_masradcal} for Galactic PNe.
Such relation offers also a method to estimate the distance of PNe,
once the ionised mass and the angular radius can be measured
(Maciel \& Pottasch 1980; see also Kwok 1985).
The mass--radius relationship is naturally 
explained in the framework of the interacting-winds model
(VK85), which couples the dynamical expansion of the nebula with
the ionising properties of the central star.

In this study we use the observed relation first to derive some 
calibration of $V_{\rm AGB}^{\rm max}$, for which we presented three 
possible prescriptions in Sect.~\ref{sssec_vagbm}.
To this aim, we calculate PN models all corresponding to a CSPN
with mass of about 0.6 $M_{\odot}$, since this should be
representative of the typical PN population in the Galactic disk
(see Stasi\'nska et al. 1997, and references therein). 
The resulting values for $V_{\rm AGB}^{\rm max}$ are: 
i) $const. = 15$ km s$^{-1}$ (models 1 and 5), 
ii) $F(Z) \sim 27$ km s$^{-1}$ (model 2)
according to Bressan et al. (1998), with $Z$ being the effective metal
abundance (higher than the initial value because of dredge-up
episodes); iii) 
$F(L) \sim 18$ km s$^{-1}$ (model 3) according to  Kr\"uger et al. (1994). 
We can notice that all the corresponding curves in Fig.~\ref{fig_masradcal}
lie below the mean line of the observed data, even when  a shorter
transition time ($t_{\rm tr} = 500$ yr, model 5) is adopted
to reduce the discrepancy. 

We find that the mean mass-radius relation can be well 
reproduced by decreasing $V_{\rm AGB}^{\rm
max}$, e.g. applying a multiplicative factor $f_{\rm v} = 0.588$ to
the original  Kr\"uger et al. data (so that $V_{\rm AGB}^{\rm max} 
\sim 10$ km s$^{-1}$ in model 4 ; see also Fig.~\ref{fig_vagbm}).
Most of our PN model calculations,  
presented in the next sections,  are carried out by adopting this 
calibrated prescription.
An additional advantage, compared to the case of 
a constant velocity, is that such a choice 
produces a natural dispersion of the expansion velocities at the end of
AGB,  which seems more consistent with the observed scatter of the expansion
velocities of young PNe at small radii (see Fig.~\ref{fig_velrad}).

\begin{figure*}
\begin{minipage}{0.69\textwidth}
\resizebox{\hsize}{!}{\includegraphics{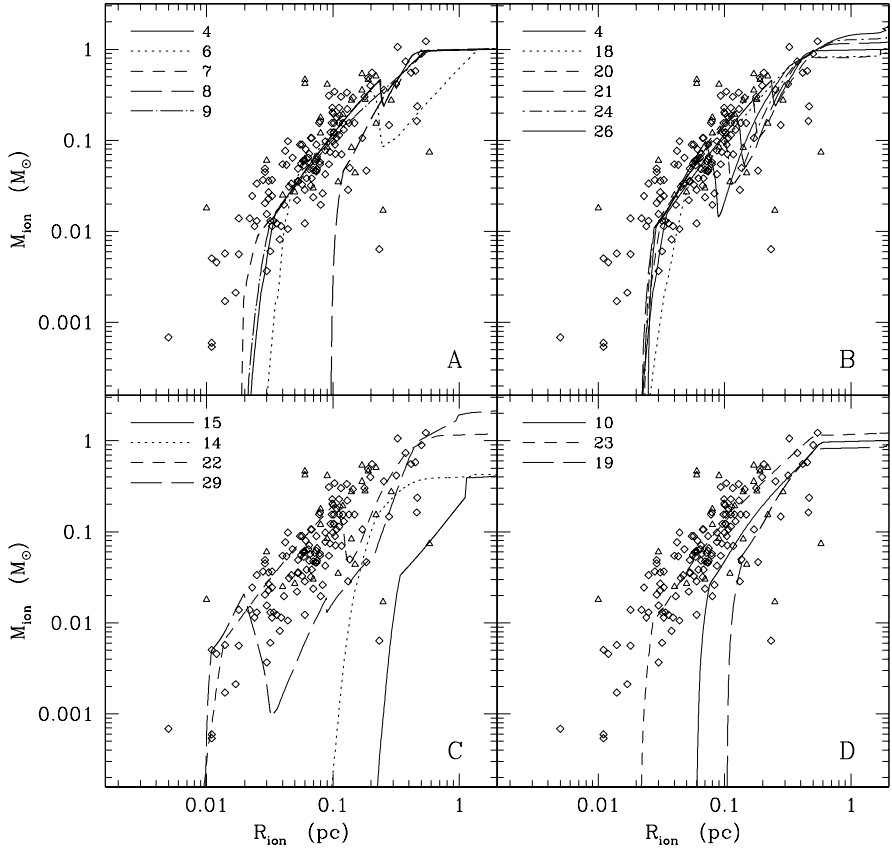}}
\end{minipage}
\hfill
\begin{minipage}{0.30\textwidth}
\caption{Ionised mass -- nebular radius relationship.
Observed data are the same as in Fig.~\protect{\ref{fig_masradcal}}.
Theoretical predictions are shown for several choices of the
parameters (see Table~\protect{\ref{tab_mod}})
}
\label{fig_masrad}
\end{minipage}
\end{figure*}

With this calibration for $V_{\rm AGB}^{\rm max}$, we show
in Fig.~\ref{fig_masrad} the predictions for several
choices of the other model parameters (quoted in Table~\ref{tab_mod}).
In general, all tracks exhibit a common feature
in the $M_{\rm ion}$--$R_{\rm ion}$ plane, i.e. a change
in the slope after an initial steep 
increase of the ionised mass. This point marks the transition
from ionisation- to density-bounded shell configurations.
In other words, during the early epochs 
the ionisation front proceeds forward but keeping behind 
the outer shock, so that the nebula is 
optically thick to the ionising photons, and 
the outermost part of the shocked shell is still neutral.

As soon as the ionisation front reaches the outer rim of the shell, 
the unperturbed (low-density) AGB wind is also ionised and the nebula
becomes optically thin to the continuum photons emitted by the central star.
During these stages the whole shell is ionised, and the 
theoretical mass-radius tracks merely represent the evolution 
of the shell parameters. The almost linear behaviour 
proceeds until when --  due to the decrease of the ionising flux
from the central star --  the ionised mass starts to decrease as well,
and correspondingly the ionisation front recedes inside the shell.
At later ages, $M_{\rm ion}$ increases again as the nebula expands and its
density decreases. Finally, when the AGB wind is entirely engulfed by 
the shocked shell, $M_{\rm ion}$ remains constant at increasing 
radius (horizontal part of the tracks).   

\begin{figure}
\resizebox{\hsize}{!}{\includegraphics{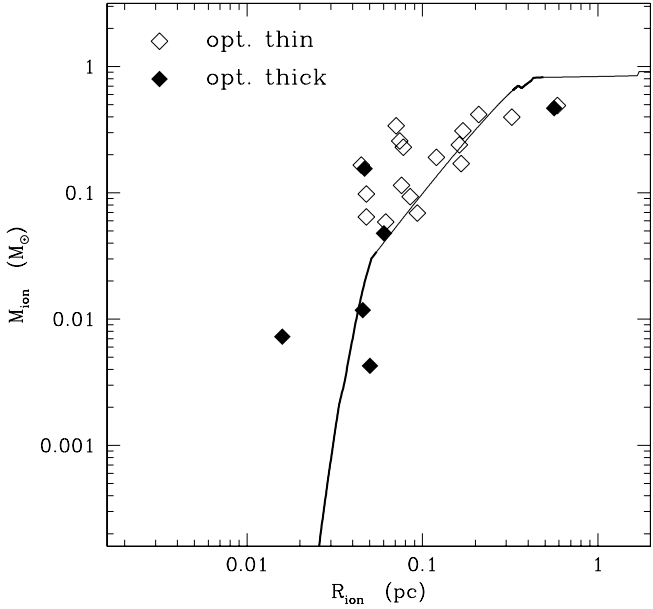}}
\caption{Ionised mass as a function of the nebular radius 
for the sample of Galactic PNe analysed by M\'endez et al. (1992).
Optically-thin/thick sources are distinguished by means of
empty/filled symbols, respectively.
Along the theoretical track, corresponding to model 18, the 
expected optical properties are indicated (thick line: 
optically-thick configuration with $\mu = 1$;  thin line: 
optically-thin configurations with $\mu < 1\,$)}
\label{fig_masradtt}
\end{figure}

The evolution of the optical properties of PNe along a mass--radius track
is also shown in Fig.~\ref{fig_masradtt}, together with the observed data
for the sample of Galactic PNe studied by M\'endez et al. (1992),
who estimate the $\mu$-factor characterising each object.
Both the location and the slope of the theoretical relationship for
optically thick/thin sources are in excellent agreement with the
observed data. 

Clearly, this general trend may be affected by various parameters. 
Let us first discuss the models  
illustrated in panel A of Fig.~\ref{fig_masrad}, 
all corresponding to a central star with mass
of $\approx 0.6 M_{\odot}$. 
Most of these PNe models account for the mean observed 
mass-radius relation quite well, which is an expected result
as they should be representative of the typical Galactic PNe.
For a given transition time ($t_{\rm tr} = 1500$ yr), the largest differences
show up at the lowest and largest values of $M_{\rm ion}$, whereas
the models almost converge during the optically-thin stages.
For a significantly longer $t_{\rm tr}$ (8000 yr) the observed data
are intercepted by the theoretical track (model 8) only at the largest 
$R_{\rm ion}$, since the nebula starts to be ionised when it has already 
expanded to relatively large dimensions.

The dependence on the mass of the central star can be visualised
in panel B of Fig.~\ref{fig_masrad}.
It turns out that the radius at which ionisation first sets in 
does not vary much with $M_{\rm CSPN}$, being essentially determined
by the transition time (the same in all models).
A clear effect of $M_{\rm CSPN}$ is instead related to the part of the tracks
following the transition from optically-thin to optically-thick
configurations (the first relative maximum in the curves).
At increasing stellar mass, this transition occurs at shorter radii
reflecting the faster evolutionary speed. 
Moreover, we can note that the non-monotonic behaviour of the tracks
can explain part of the observed scatter of $M_{\rm ion}$ at given
$R_{\rm ion}$.

The effect  of the  transition time can be analysed in panel C.
In this panel, we can notice that the observed PNe 
with the lowest masses and radii would be consistent 
with relatively massive central stars and quite short transition times
($t_{\rm tr} =  500$ yr; models 22, 29). 
Models with lower $M_{\rm CSPN}$  and much longer $t_{\rm tr}$ 
(10000 yr; models 15, 14) succeed in explaining the 
few data at quite large $R_{\rm ion}$ and relatively low $M_{\rm ion}$.   

Finally, we plot in panel D  the expected mass-radius tracks 
for He-burning post-AGB tracks.  
These models are consistent with the observed data as well 
as those with H-burning central stars. 
Comparing models 21 (H-burner) and 23 (He-burner) --  
which are calculated with the same set of the other parameters --
we can see that, contrary to the H-burning case, 
in the He-burning model the ionised mass always
increases with radius and the nebular shell, once becomes optically thin, 
remains density-bounded up to the largest radii.
This can be explained considering that He-burning tracks are, 
on average,  characterised 
by quite longer evolutionary time-scales, so that the drop of the ionising
flux emitted by the central star occurs when the expanding shell has already 
reached large radii.

\begin{figure}
\resizebox{\hsize}{!}{\includegraphics{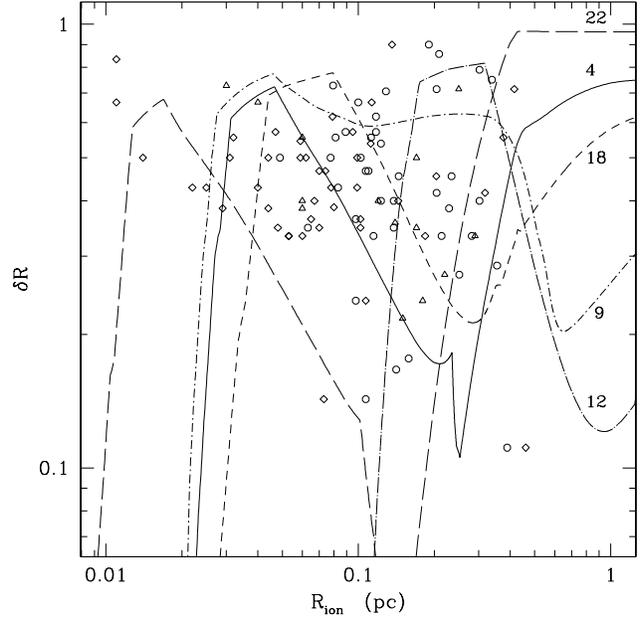}}
\caption{Relative shell thickness as a function of
the nebular radius. Measured radii are taken from  Boffi \& Stanghellini 
(1994; triangles); Zhang (1995; squares); and 
G\'orny et al. (1997; circles). Corresponding semi-empirical determinations
of $\delta R$ are those derived by Zhang \& Kwok (1998).
A few theoretical predictions are shown for comparison
(see Table~\protect{\ref{tab_mod}})}
\label{fig_thick}
\end{figure}

\subsection{The nebular shell thickness}
PNe exhibit a large variety of morphological and geometrical features
(e.g. Schwarz et al. 1992),
that clearly cannot be handled by models that assume spherical symmetry.
Nevertheless, we can at least test our predictions
concerning  an important geometrical  parameter, 
namely the relative shell thickness
$\delta R$ (defined by Eq.~\ref{eq_dr}).
As already described in Sect.~\ref{ssec_thick}, our model considers  
the effect of ionisation in determining the thickness of the shell.
    
To construct Fig.~\ref{fig_thick}, we combine determinations of nebular
radii taken from different sources with the recent semi-empirical 
determinations of $\delta R$ (for the primary shell) 
obtained by Zhang \& Kwok (1998) 
in their morphological  analysis of a sample of Galactic PNe. For each of
them, Zhang \& Kwok (1998) derive relevant morphological parameters,
by constraining simulated optical and radio images to reproduce
the observed ones. It turns out that $\delta R$ mostly varies over the range
$[0.1, 0.8]$, with a  distribution peaked at around 0.45.

We can see that the theoretical predictions (a discussion of the 
trends is given in Sect. \ref{ssec_thick})
account well for the location 
of the observed data in the $\delta R$--$R_{\rm ion}$ plane.
Together with the good reproduction of the ionised mass--radius 
relationship, this fact implies that both the dynamical and 
ionisation properties of PNe should
be reasonably well represented in our model.

\subsection{Electron densities}
%

\begin{figure}
\resizebox{\hsize}{!}{\includegraphics{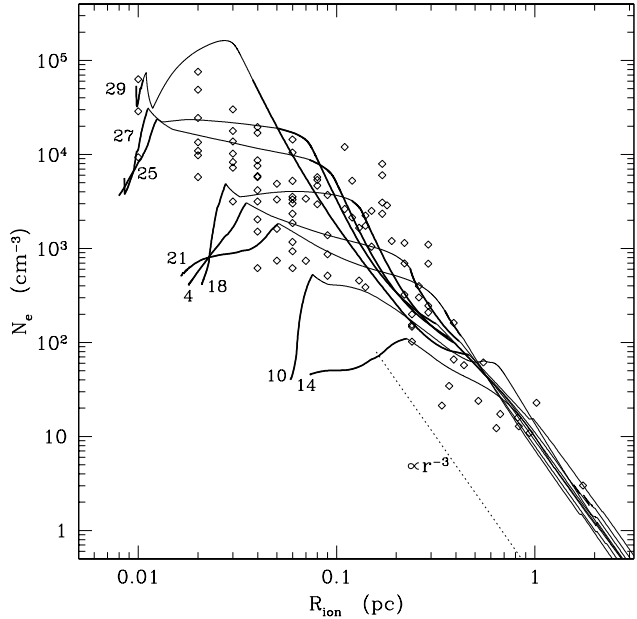}}
\caption{Electron densities as a function of the nebular size.
Observed data for Galactic PNe are taken from the compilation 
by Phillips (1998).
Model predictions are shown for comparison. The straight line,
$N_{\rm e} \propto R_{\rm ion}^{-3}$, is plotted to indicate the
expected slope for constant nebular mass}
\label{fig_radnel}
\end{figure}

Figure \ref{fig_radnel} displays the radial variation of the  
nebular electron densities.
Empirical determinations of $N_{\rm e}$ are all 
derived from the [S{\sc \,ii}]$\,\lambda6717/6730$ forbidden line ratio,
which should sample the outer nebular regions.
In this way, we expect that the considered PNe sample is little subject to 
the effects of radial stratification
in ionisation and density -- present when different lines are used --
as well as not dependent on the nebula filling factor, 
which is the case when $N_{\rm e}$ is estimated 
from the radio continuum/H$\beta$ fluxes.

{From} the observed distribution of the data points    
it should be noticed that, first of all, the relation  between 
$N_{\rm e}$ and  $R_{\rm ion}$ shows a relatively large dispersion, which
seems to reduce towards larger radii.
Second, as pointed out by Phillips (1998), there is clear evidence 
for a change in the mean slope of the radial gradient at 
$R_{\rm ion} \approx  0.1$ pc.
Both features are quite well reproduced by our models, showing
the evolution of the average density of the ionised shell as function
of the radial size of its outer boundary.
The data points located at $R_{\rm ion} <  0.1$ pc would be consistent
with a mixture of both optically-thick and thin nebulae, the latter 
being characterised by a more gradual decrease of  $N_{\rm e}$ with 
radius. Indeed, this feature would explain the observed scatter of the points
at smaller $R_{\rm ion}$.
For $R_{\rm ion} >  0.1$, the reduced dispersion of the observed data 
which, on average, seem to follow the 
relation $N_{\rm e} \propto R_{\rm ion}^{-3}$, 
could be interpreted by the gradual convergence of all 
PNe to the condition  $M_{\rm ion} = M_{\rm sh} = \Delta M_{\rm AGB}$, i.e.
the shocked ionised shell embraces all the AGB ejecta, so that its
mass remains constant.  
Finally, we note that the theoretical  $N_{\rm e}-R_{\rm ion}$ tracks 
show a systematic trend with the mass of the central star, i.e. 
at smaller radii the largest densities are attained for more 
massive CSPN. However, it should be remarked that the predicted $N_{\rm e}$
refer to average electron densities, which do not account for intrinsic
density gradients inside the nebulae.

\subsection{PN central stars: temperatures and luminosities}
\label{ssec_zan}
%

\begin{figure}
\resizebox{\hsize}{!}{\includegraphics{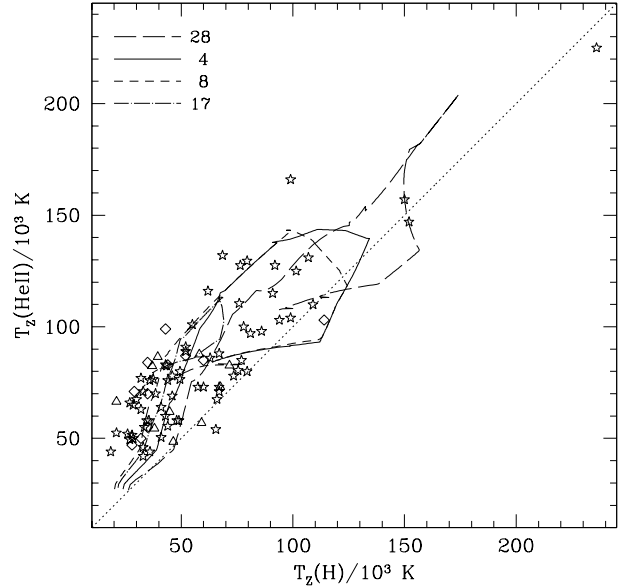}}
\caption{Zanstra discrepancy between $T_{\rm Z}$(H) and 
$T_{\rm Z}$(He{\sc \,ii}).
Observed data for Galactic PNe 
are taken from de Freitas Pacheco et al. (1986; 
triangles); Gleizes et al. (1989; starred symbols); and M\'endez et al.
(1992; squares). Model predictions for different choices of the 
parameters are shown for comparison}
\label{fig_zandisc}
\end{figure}

Determining  
the main parameters of the central stars of PNe , i.e. effective temperature
and luminosity, means to cast light on important aspects of both 
nebular and stellar evolution. For instance, 
by assigning a position on the H-R diagram to the ionising star,  
it is possible to derive the stellar mass 
-- from the comparison with theoretical post-AGB tracks --  
and estimate (an lower limit to) the PN dynamical 
age, if the expansion velocity and nebular sizes are measured.
The combined information on stellar mass and PN age, 
offers also the possibility
to discriminate between H- and He-burning stars (e.g. Dopita et al. 1996).  
 
\begin{figure*}
\begin{minipage}{0.69\textwidth}
\resizebox{\hsize}{!}{\includegraphics{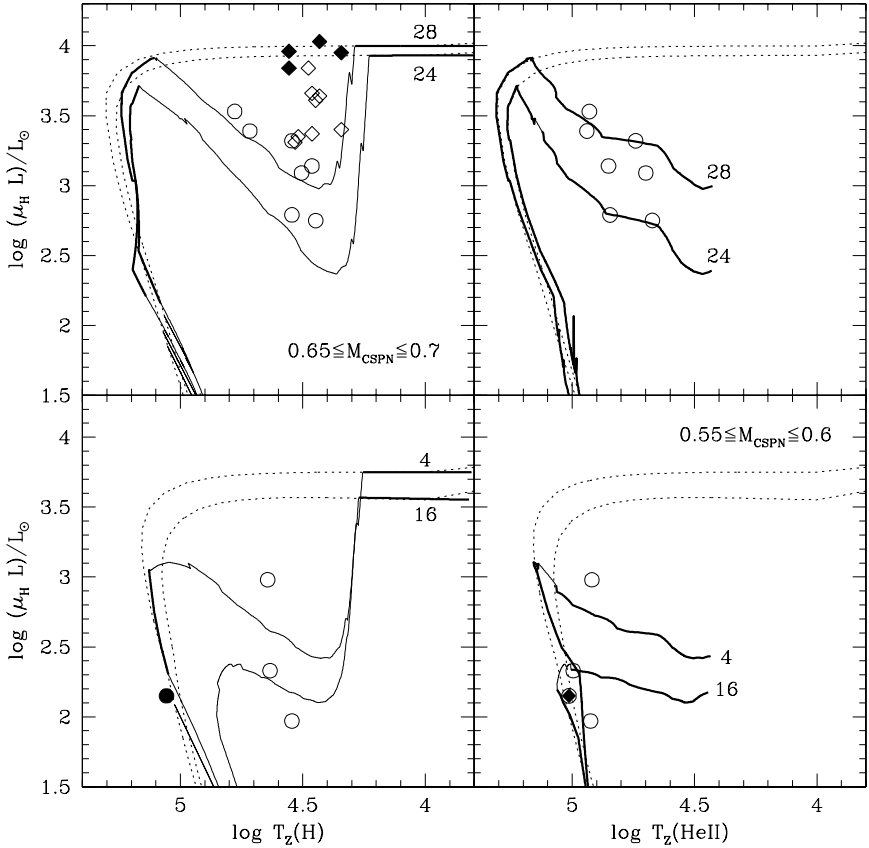}}
\end{minipage}
\hfill
\begin{minipage}{0.30\textwidth}
\caption{Zanstra temperatures 
($T_{\rm Z}({\rm H})$ in the left panels 
and $T_{\rm Z}({\rm He\mbox{\sc \,ii}})$ in the right ones) 
and luminosities ($\mu_{\rm H} L$) of PN central stars.
The upper panels present the nebulae with CSPN masses 
$0.65\le M_{\rm CSPN}\le0.7$, whereas the lower panels refer
to those with $0.55\le M_{\rm CSPN}\le0.6$.  
Observed data points are taken from M\'endez et al. (1992), and 
shown with filled squares for $\mu_{\rm H} > 0.7$ and empty squares
for  $\mu_{\rm H} < 0.7$. PNe of low-excitation (with $T_{\rm Z}({\rm H}$
only) and high-excitation (with both $T_{\rm Z}({\rm H}$ and 
$T_{\rm Z}({\rm He\mbox{\sc \,ii}})$) are denoted with squares and 
circles, respectively.
Model predictions are superimposed distinguishing 
optically-thin/thick configurations
(thin/thick solid lines) in the H and He$^{+}$ continua, together with
the corresponding $\log L - \log T_{\rm eff}$ tracks (dotted line).
The masses of the central stars are chosen consistently with those
estimated by M\'endez et al. (1992) for the sample PNe
		}
\label{fig_hrz}
\end{minipage}
\end{figure*}

Unfortunately, the determination of luminosities of 
Galactic CSPN is indeed problematic,
mostly due to the large uncertainties that often affect the observational 
methods to estimate effective temperatures and distances.
The effective temperature of a CSPN can be determined either directly, 
by fitting observed spectra with model atmospheres,  or   
with the aid of indirect methods, such as the Zanstra and 
energy-balance method
(see, e.g.,  Gathier \& Pottasch 1989). 
The former, first introduced by Zanstra (1931), is based on the 
the assumption that the nebula is optically thick to 
the ionising continuum under consideration 
-- the H and He{\sc \,ii} Zanstra temperatures 
$T_{\rm Z}$(H) and $T_{\rm Z}$(He{\sc \,ii}) are usually determined -- 
and requires
the measurement of both nebular and central star optical emissions.
The latter, first discussed by Stoy (1933), is based on the assumption
that the ratio of the total energy emitted by a nebula in forbidden lines 
to the energy emitted in one of the hydrogen recombination lines depends on
the temperature of the central star.
Then, once $T_{\rm eff}$ is estimated, 
the luminosity of the CSPN can be derived, for instance,  
i) from the apparent magnitude at a given wavelength provided
that the distance is known, ii) from the comparison with post-AGB tracks 
in the $\log g - \log T_{\rm eff}$ diagram if estimates of the 
surface gravity $g$ and mass of the CSPN are available; 
iii) by equating the nebular emission over the whole spectral range 
to the radiation emission from the central star, 
with the assumption that the nebula is ionisation-bounded
(see, e.g. Gathier \& Pottasch 1989, M\'endez et al. 1992 for 
description and application of the methods).

The so-called ``Zanstra discrepancy'' between 
$T_{\rm Z}$(H) and $T_{\rm Z}$(He{\sc \,ii}) --
i.e. usually  $T_{\rm Z}$(H) $<$  $T_{\rm Z}$(He{\sc \,ii}) -- 
has been a matter
of debate for long time (see, e.g., Stasi\'nska \& Tylenda 1986; 
Gathier \& Pottasch 1989; M\'endez et al. 1992; Gruenwald 
\& Viegas 2000 and references therein).
The discussion aims at identifying the main causes of such discrepancy 
and their role, namely: 
i) the condition of optical thinness to the H continuum;
ii) the possible excess of photons able to ionise He$^+$ 
-- with energy $h\nu>54.4$~eV -- emitted 
by the CSPN,  compared to the black-body spectral energy
distribution; 
iii) the differential dust absorption in the nebula.    

Figure~\ref{fig_zandisc} illustrates the Zanstra discrepancy for Galactic PNe.
Model predictions for $T_{\rm Z}$(H) and $T_{\rm Z}$(He{\sc \,ii}) 
are derived from the conditions:
	\beqa
	\mu_{\rm H} Q({\rm H}^0)
	& = & 4 \pi \int_{13.6}^{54.4}
	\frac{B_{\nu}[T_{\rm Z}({\rm H})] }{h\nu} {\rm d}(h\nu) 
	\label{eq_zh} \\
	\mu_{{\rm He}^{+}} Q({\rm He}^+)
	& = & 4 \pi \int_{54.4}^{\infty}
	\frac{B_{\nu}[T_{\rm Z}({\rm He\mbox{\sc \,ii}})] }{h\nu} {\rm d}(h\nu)
	\label{eq_zhe} 
	\eeqa
where $B_{\nu}$ is the Planck function and the 
other quantities have the usual meaning. It should be noted that we do not  
strictly follow the same procedure adopted by observers to obtain the 
Zanstra temperatures, which usually involves
the ratio between nebular and stellar emission at particular wavelengths
(the so-called Zanstra ratios; see e.g. M\'endez et al. 1992). However,
Eqs.~(\ref{eq_zh}) and (\ref{eq_zhe}) share the same basic assumptions, 
i.e. the nebula is considered in equilibrium state 
between ionisation and recombination processes, and optically thick 
to the ionising continuum (ionisation-bounded).
 
The theoretical tracks in Fig.~\ref{fig_zandisc} account for the observed
trend of the Zanstra discrepancy. 
{From} the analysis of the models it is worth    
reporting that i) at lower $T_{\rm Z}$(H) the Zanstra discrepancy increases
with both the stellar mass and the transition time, and   
ii) the highest values of $T_{\rm Z}$(He{\sc \,ii}) are attained
by the most massive models; iii) there is some hint for an inverse Zanstra 
discrepancy (i.e. $T_{\rm Z}$(H)$ > T_{\rm Z}$(He{\sc \,ii})) which may 
occasionally occur when the central star overcomes the hottest 
knee of its post-AGB track and starts to decline in luminosity.

Figure \ref{fig_hrz} contributes to investigate the causes of the Zanstra 
discrepancy. To this aim, we adopt the sample of 23 Galactic PNe
analysed by M\'endez et al. (1992), for which a large number of parameters are 
determined, including: the nebular masses and radii 
(see also Sect.~\ref{sssec_masrad}), the masses of the CSPN,
the effective temperatures, the Zanstra temperatures, etc.  
The observed data for  $T_{\rm Z}$(H) and $T_{\rm Z}$(He{\sc \,ii}) are plotted 
as a function of the quantity $y L_a$ as defined by M\'endez et al. (1992;
see their table 4), which is equal to $\mu_{\rm H} L$ according to their
equation~12. The comparison with the theoretical predictions is 
then made for representative models whose stellar masses fall within the range
indicated by M\'endez et al. 

The general accordance between the observed data and their expected location
is remarkably good. 
The observed dispersion of the data  
in the $\log(\mu_{\rm H} L) - \log T_{\rm Z}$(H) plane (see top-left panel
of Fig.~\ref{fig_hrz}) is naturally accounted for by the models, which 
exhibit notable deviations from the horizontal parts (at constant luminosity) 
of the $\log L - \log T_{\rm eff}$ post-AGB tracks.
Such deviations show up as soon as the ionised shell becomes 
optically thin (i.e. $\mu_{\rm H} < 1$) to the H continuum, 
which confirms the conclusions already drawn by M\'endez et al. (1992).
Note, however, that there is some offset in $T_{\rm eff}$ 
between observations 
and models, i.e. the latters get optically thin at lower effective
temperatures than indicated by the data.

In the  $ \log( \mu_{\rm H} L) - \log T_{\rm Z}$(He{\sc \,ii}) plane 
(see top-right panel of Fig.~\ref{fig_hrz})
the models succeed  in reproducing the 
corresponding data which, in this case,  are 
expected to correspond to optically thick configurations 
to the He$^+$ continuum. 
It is also worth noticing that the agreement holds as well when the data
are analysed as a function of the mass of the CSPN, as is clear
by comparing the top panels (higher masses) to the bottom ones
(lower masses) of Fig.~\ref{fig_hrz}.
Finally, a general convergence is found with respect to the discrimination
between optically thin/thick nebulae and their location in the
observational planes.

\subsection{Expansion velocities}
%
\begin{figure*}
\begin{minipage}{0.69\textwidth}
\resizebox{\hsize}{!}{\includegraphics{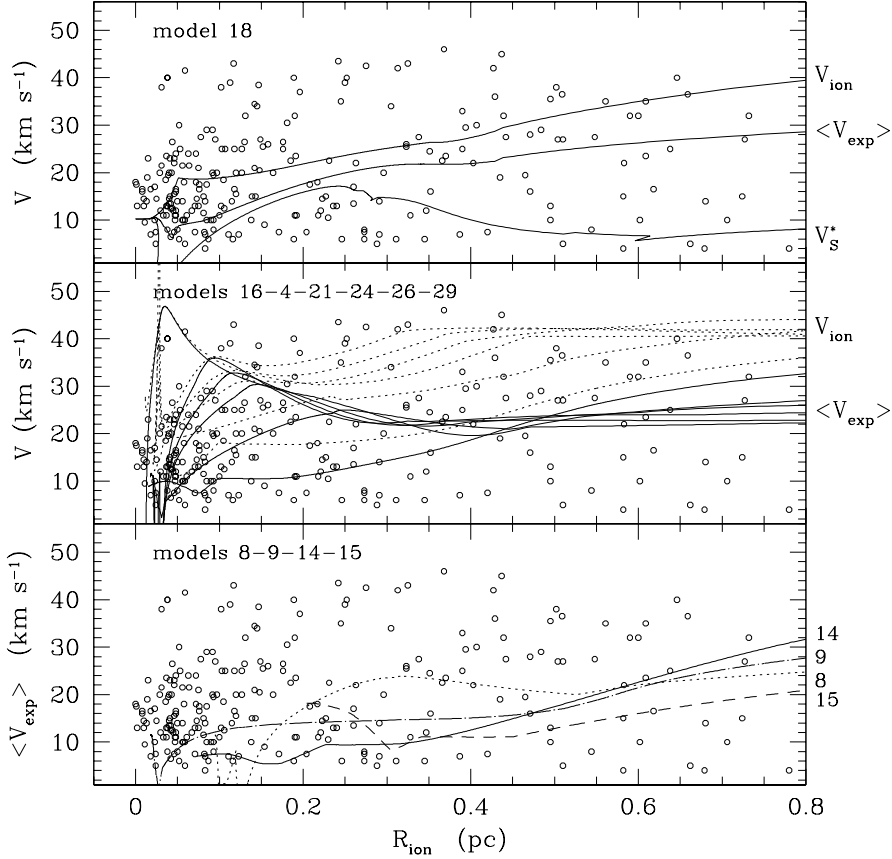}}
\end{minipage}
\hfill
\begin{minipage}{0.30\textwidth}
\caption{Expansion velocities as a function of the nebular
size of Galactic PNe. Data (circles) derived from the
forbidden lines of [O{\sc \,iii}] and [O{\sc \,ii}]  
are taken from the compilation by Weinberger (1989).
Model predictions are shown for comparison for several choices
of the parameters. See text for further details }
\label{fig_velrad}
\end{minipage}
\end{figure*}
The expansion velocities of  PNe -- 
usually derived from  emission lines like  H{\sc \,i}, He{\sc \,ii},
[S{\sc \,ii}], [O{\sc \,ii}], [Ar{\sc \,iv}], [Ne{\sc \,iii}], 
and more often [O{\sc \,iii}] -- 
are found to range from few to 40 -- 45 km s$^{-1}$, without
any well-defined correlation with the nebular size 
(see Fig.~\ref{fig_velrad}). 

Moreover, velocity estimates based on lines produced by ions with
different ionisation potentials, have revealed the existence of  velocity 
gradients across the nebulae (Gesicki et al. 1998). 
In general, the velocity is found to increase with the ionisation
potential of the emitting ion, i.e. going from the inner to the 
outermost layers of the nebula.
The fact that a large scatter (shown in Fig.~\ref{fig_velrad}) 
still characterises the velocity data,  though
derived from the same lines, likely points to a relatively broad 
range of parameters affecting
the dynamics of PNe.

We attempt to investigate this point on the basis of the analytical 
prescriptions already presented in Sects.~\ref{ssec_dyn} and \ref{ssec_thick}.
We just recall that, although the adopted dynamical description is a
simplified approach that cannot account for the complex velocity structure
shown by hydrodynamical simulations 
(e.g. Schmidt-Voigt \& K\"oppen 1987b; 
Marten \& Sch\"onberner 1991), 
nevertheless it should provide a reasonable representation of 
the average dynamical properties of the expanding nebulae. 
Furthermore, a clear improvement over similar analytical models is that
the expansion of the nebula is not expressed with a single velocity, 
but  the velocities of both the inner and outer rims of the ionised shell
($V_{\rm S}^*$, and $V_{\rm ion}$ respectively) are predicted.
In this way, the analytical model 
keeps track, to some extent, of the real velocity gradient inside the nebula.

An example is shown in  Fig.~\ref{fig_velrad} (top panel), where
the two rim velocities are plotted as a function of $R_{\rm ion}$.
We can immediately notice the 
relatively large velocity range 
comprised by the two curves. In particular, quite low velocities 
(few km s$^{-1}$) are attained by the contact discontinuity during the 
early post-AGB stages, when the increased pressure
due by ionisation slows down its expansion (see Sect.~\ref{ssec_thick} for 
more discussion). Conversely, the ionisation front is always accelerating 
outward.

On the other hand, 
for purpose of easier comparison with observed data, 
we may also define an average expansion velocity 
	\beq
	\langle V_{\rm exp}\rangle = 
\frac{3}{4 \pi (R_{\rm ion}^3-R_{\rm S}^{* 3})} 
	\int_{R_{\rm S}^*}^{R_{\rm ion}} V(r) \, 4 \pi r^2 {\rm d} r 
	\eeq
following Schmidt-Voigt \& K\"oppen (1987b). In practice, it is a
weighted mean over the nebular (emitting) volume,  under the assumption
that the velocity varies linearly between the two extrema 
$V_{\rm S}^*$ and $V_{\rm ion}$.
According to this definition $\langle V_{\rm exp}\rangle$ 
is closer to  $V_{\rm ion}$,
that is the maximum velocity.

The dependence of $\langle V_{\rm exp}\rangle$ on the 
CSPN mass (for H-burning models) is evident already during the initial
stages of the nebular expansion, 
so that the highest velocities are reached by the 
most massive models (see middle panel of Fig.~\ref{fig_velrad}).
We can also notice that all the rising branches of the curves 
concentrate on a 
narrow range in radius, just reflecting the fact that the transition times 
of  these models are comparable (i.e. 1500 yr in most cases).
After the initial rise and peak, the curves start then to decline and finally
flatten out at velocities of about  20 -- 30 km s$^{-1}$, which coincides 
with the mean range of the observed data. However,  
we remark again that this is a just a  mean trend. 
For the same set of models the velocities   
of the outer rims are shown for comparison (dotted lines).

The initial radius of each velocity curve essentially represents the stage 
at which the nebula starts being ionised by the central star. This radius, 
$R_{{\rm ion},0}$, can be expressed in first approximation as 
$ V_{\rm AGB}^{\rm max} \times t_{\rm tr}$,
i.e. the product of  the velocity of the ejected AGB material and the 
transition time.
At increasing transition times, ionisation sets in at larger radii (compare
model 4 with 8, having $t_{\rm tr} = $ 1500 and 8000 yr, respectively). 
The same effect is produced for lower values of the ejection velocity, as
we can see in the bottom panel 
by comparing  model 14 (solid line) with model 15 
(short-dashed line), having both $t_{\rm tr} = 10\,000$ yr, but
$V_{\rm AGB}^{\rm max} \sim$ 6, 17  km s$^{-1}$ respectively.
It follows that at least a part 
of the observed velocity spread of PNe at
any radius can reflect differences in the evolutionary speed of the central 
star, as well as be the result of differences in the ejection velocity 
on the AGB.

It is worth also noticing  the different behaviour of 
$\langle V_{\rm exp}\rangle$
of PNe ionised by central stars  that are either H- (e.g. model 4) 
or He-burners (e.g. model 9). On average, we expect that at given nebular
size a PN ionised by a He-burning CS has a lower expansion velocity than
that surrounding a H-burning CS of the same mass.  
This in turn would translate into longer dynamical ages, 
-- defined as $t_{\rm dyn} = R_{\rm ion}/V_{\rm exp}$ --  
of PNe with He-burning CSs.
This age difference offers a viable way to discriminate
between the two classes of stars (see Dopita et al. 1996 for 
an application to PNe in the LMC). 

On the basis of all mentioned points it is clear that, even if we 
are interested in average dynamical properties of PNe,  
the often adopted assumption of a constant expansion velocity can lead to 
substantial errors in the estimation of the true dynamical ages.
Actually, it will be interesting  
to use our model results in a future analysis 
to address the 
discrepancy between the evolutionary and dynamical ages of PNe
(see e.g. McCarthy et al. 1990).

\begin{figure*} 
\resizebox{0.9\hsize}{!}{\includegraphics{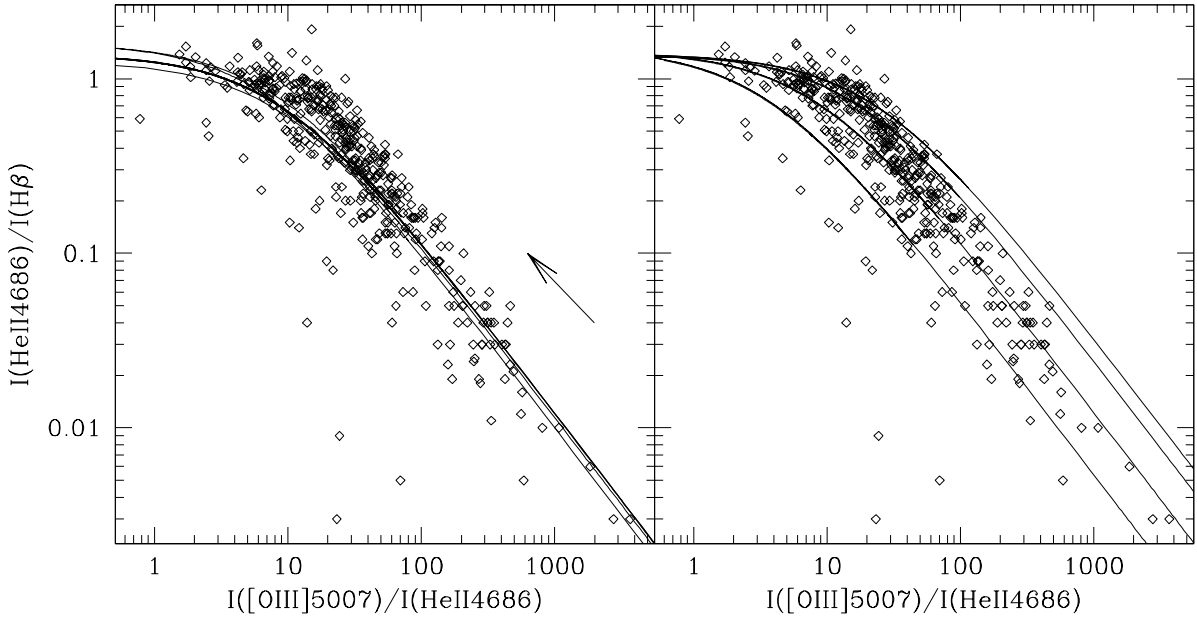}}
\caption{Nebular line ratios in Galactic PNe. Data are taken from the
Strasbourg-ESO Catalogue of Galactic Planetary Nebulae (Acker et al. 1992).
Theoretical tracks are superimposed for comparison.
Let panel: models 4--8--9--14--28, all calculated assuming a constant
electron temperature $T_{\rm e} = 10\,000$ K.
Right panel: models 4--11--12--13, calculated with $T_{\rm e} = 
8\,000$, $10\,000$, $12\,000$ and $13\,000$ K (from higher to lower
tracks). The arrow indicates the direction of increasing $T_{\rm eff}$ 
}
\label{fig_o3he2} 
\end{figure*}

\subsection{Nebular line ratios}
\label{sec_lineratios}
The ionisation structure and line emissivities are among the most
basic properties of our PNe models, and can be easily checked
by comparison with observations.

\subsubsection{The ${\rm He}\mbox{\sc \,ii}/{\rm H}\beta$ 
vs. $[{\rm O}\mbox{\sc \,iii}]/{\rm He}\mbox{\sc \,ii}$ 
	anticorrelation}

Let us start with Figure~\ref{fig_o3he2}, 
which combines the intensities of the three 
optical emission lines here considered, i.e. recombination 
(H$\beta$, He{\sc \,ii}$\,\lambda4686$) and forbidden 
([O{\sc \,iii}]$\,\lambda5007$) lines.

First of all, we notice that the observed data do not present a 
large scatter, but rather follow a well-defined  relation.
There is an evident anticorrelation between 
the He{\sc \,ii}$\,\lambda4686$/H$\beta$ 
and [O{\sc \,iii}]$\,\lambda5007$/He{\sc \,ii}$\,\lambda4686$ 
intensity line ratios, 
until the former becomes almost
constant ($\ga 1$) for low values of 
[O{\sc \,iii}]$\,\lambda5007$/He{\sc \,ii}$\,\lambda4686$.
 
The interpretation of such trend can be easily derived when considering that
 the ionisation potentials of He$^+$ (54.4 eV) and O$^{++}$ (54.9 eV)
are almost identical. It follows that the Str\"omgren volumes 
$V$(He$^{+}$) and $V$(O$^{++}$) nearly coincides and the 
lines He{\sc \,ii}$\,\lambda4686$ and [O{\sc \,iii}]$\,\lambda5007$ 
are mainly produced within separate
nebular regions, whose complementarity in  emission volumes  
(i.e. when the innermost 
$V$(He$^{++}$) increases, the outer $V$(O$^{++}$) decreases and vice-versa)  
explains the observed anticorrelation between the respective line 
intensities.

It is also clear that since He$^{++}$ is present in the nebula only
if the central star is sufficiently hot, the observed trend  in
Fig.~\ref{fig_o3he2} implicitly contains a temperature scale
(increasing along the direction of the arrow), and as a matter of fact,
it has been used as a semi-empirical method to determine the 
temperature of high-excitation CSPN (see e.g. Gurzadyan 1988). 

It is really striking how the empirical relation is so well reproduced
by our models, which stands on few simple prescriptions for the
ionisation and emission nebular structure (see Sect.~\ref{ssec_lines}).
In this respect we draw the attention to the following points:

The flattening of the tracks towards the lowest 
[O{\sc \,iii}]$\,\lambda5007$/He{\sc \,ii}$\,\lambda4686$
values occurs when the nebulae become optically-thin
to both the H and He$^{+}$ continua. In fact, as the
emission volumes for these ions are the same,  
the  He{\sc \,ii}$\,\lambda4686$/H$\beta$
intensity ratio depends only on atomic parameters
(i.e. recombination coefficients and chemical abundances).

Finally, comparing the two panels of Fig.~\ref{fig_o3he2}, it is evident
that for given  He{\sc \,ii}$\,\lambda4686$/H$\beta$
intensity ratio, the relative intensity 
of the forbidden line is  a strong function of the electron temperature
(right-panel), whereas it is almost independent of other parameters
(e.g. transition time, mass of the CSPN, H-/He-burning conditions, etc.).
As expected, most of the observed data can be reproduced if $T_{\rm e}$
is varied from about $8\,000$ to $13\,000$ K, in agreement with 
the typical temperature range indicated by observations.

\subsubsection{The evolution of the 
$I({\rm He}\mbox{\sc \,ii}\,\lambda4686)/I({\rm H}\beta)$ 
line intensity ratio }

\begin{figure} 
\resizebox{\hsize}{!}{\includegraphics{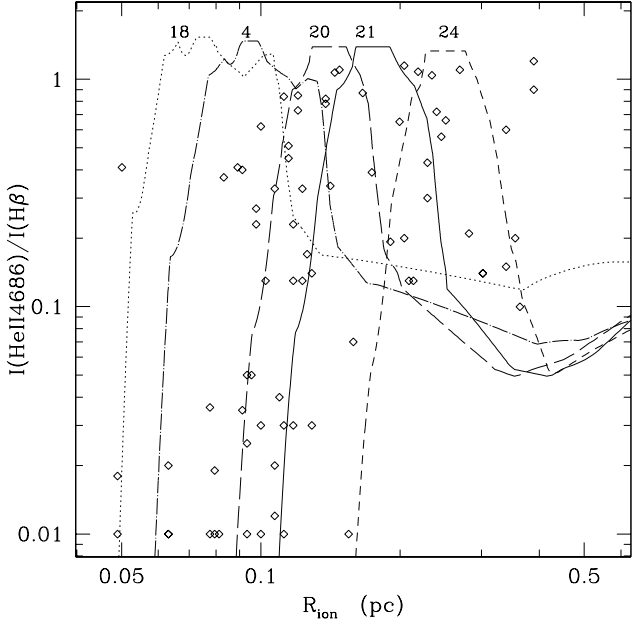}}
\caption{Intensity line ratio as a function of the nebular
size. Observed data refer to the sample of Galactic PNe 
analysed by G\'orny et al. (1997).
Model predictions are shown for different choices of
the CSPN mass
}
\label{fig_hehbeta} 
\end{figure}

Figure~\ref{fig_hehbeta} shows the behaviour of the 
$I({\rm He}\mbox{\sc \,ii}\,\lambda4686)/I({\rm H}\beta)$ 
intensity ratio as a function of the nebular radius.
 At smaller radii ($R_{\rm ion} \la 0.1$ pc), 
the observed data scatter over a large range, levelling up to values of $~1$.
This trend is explained by the models as the result of the progressive
increase of the stellar temperature at nearly constant luminosity, so that
for $T_{\rm eff} > 60\,000$ K helium starts to be doubly ionised
and the recombination line He{\sc \,ii}$\,\lambda4686$
 starts to be  detectable.

At this stage the nebula is already optically thin to the H-continuum 
(see Fig.~\ref{fig_mu}).
As the temperature increases, 
the Str\"omgren volume $V$(He$^{++}$) grows 
(and the He{\sc \,ii}$\,\lambda4686$ line intensity strengthens) 
until when it embraces
the whole nebula, which then becomes optically thin also to the 
He-continuum. This occurs in correspondence to the maximum temperature
attained close to the knee of the post-AGB track.
As already mentioned, this condition yields the  
$I({\rm He}\mbox{\sc \,ii}\,\lambda4686)/I({\rm H}\beta)$ 
line ratio to be practically constant and close to
unity (as shown by the horizontal part of the tracks).

As soon as the stellar temperature starts declining,  
the Str\"omgren radius $R_{\rm ion}$(He{\sc \,ii}) recedes inward, 
the He{\sc \,ii}$\,\lambda4686$ intensity weakens the H$\beta$ 
line is still relatively 
strong (and the nebula still optically thin).
We can notice that at some point in all models
the decrease of the $I({\rm He}\mbox{\sc \,ii}\,\lambda4686)/I({\rm H}\beta)$ 
line ratio slows down while the nebular
radius is still increasing. 
This happens when  the H$\beta$ line intensity  
is getting weaker as well and
the nebula returns to be optically thick to the H-continuum.
Such prediction is in perfect agreement with the observed lack of very
low $I({\rm He}\mbox{\sc \,ii}\,\lambda4686)/I({\rm H}\beta)$ 
line ratios at larger radii 
(i.e. $R_{\rm ion} > 0.2$ pc).

\section{Concluding remarks}
\label{sec_con}

In the present work we have presented in detail a new
synthetic model for PN evolution. 
On the one hand, this model is kept simple by using 
convenient approximations for most physical processes, 
and by limiting their description to analytical relations 
that can easily be integrated with time. On the other hand, 
the model improves upon previous synthetic PNe models in 
the attempt i) to consistently couple the PN phase to the 
previous AGB evolution of the stellar precursor, and ii)  
to account for dynamical effects (like that produced by ionisation) 
that are believed important to the evolution of PNe.

Although the present description considers many aspects of the PN evolution,
 it cannot, of course, be considered exhaustive.
Some approximations -- like the assumptions spherical symmetry and 
constant  electron temperature, and the somewhat simple description of 
the dynamical wind interactions -- are in fact too crude if compared with
the results of detailed hydrodynamical and photoionisation codes.
However, this is the due price
for fast computations. We remark that such synthetic 
codes probably provide the only way, at present times, to avoid the 
remarkable time requirements of hydrodynamical simulations.

This compromise is similar
to that in the field of AGB evolution
(see e.g. Groenewegen \& Marigo 2001): 
AGB stars can be described either by i) complete stellar 
evolution codes, that due to their demand of CPU-time
can be run only for few choices of stellar parameters,
or by ii) synthetic AGB codes, that treat ill-unknown quantities by means
of a parametric approach, and that can be easily run for
any choice of parameters. Synthetic AGB models, although
always recognised as crude tools, have had a crucial 
role in the understanding of many aspects of AGB evolution,
and are widely considered to be fundamental in the field.

Similarly, our synthetic PNe code is fast enough to allow the 
systematic computation of extended grids of PNe tracks -- for 
instance, for different choices of mass and metallicity distributions. 
The few free parameters (transition time, electron temperature, 
H-/He-burning tracks, etc.) can be constrained within reasonable 
limits by comparing the results with observations, 
in a way similar to the one we have presented in this paper. 
Anyway, none of these parameters is expected to be fixed 
for a given PNe population. In fact, the most promising way
to explain the wealth of PNe characteristics, is that of 
performing Monte-Carlo simulations in which the free parameters
are given reasonable probability distributions.

In the forthcoming papers, we will apply this approach to produce
synthetic populations of PNe in different galaxies. 
Main targets will be:
	\begin{itemize}
	\item
To further constrain the models by means of more detailed
comparison with the observed properties of Galactic, 
Magellanic Cloud, and other PNe populations.
	\item 
To explore the expected distribution of chemical abundances
among PNe, that may help to constrain synthetic TP-AGB models 
and their chemical yields.
	\item
To simulate the PNe luminosity function in different galaxies,
in order to test their suitability as extragalactic distance 
indicators. 
	\end{itemize}

\begin{acknowledgements}
We are grateful to K. Kifonidis and E. M\"uller for their constructive 
remarks on this work.
P.M. thanks S.R. Pottasch and J.B. Salas for useful 
discussions on PNe.
An anonymous referee is acknowledged for interesting suggestions
that could be fruitfully developed in future analyses. 
The work by L.G. has been partly financed by the Alexander
von Humboldt-Stiftung, and the TMR grant ERBFMRXCT~960086.
P.M. and L.G. also acknowledge the funding by the Italian
MURST. 

\end{acknowledgements}

\end{document}